\documentclass[aps,prd,nofootinbib,twocolumn,superscriptaddress,preprintnumbers,groupedaddress,showpacs]{revtex4-1}

\usepackage{amssymb}
\usepackage{amsmath}
\usepackage{epsfig}
\usepackage{hyperref}
\usepackage{comment}
\usepackage{color}
\usepackage{cleveref}
\usepackage{multirow}
\usepackage{capt-of}

\usepackage{graphicx}
\usepackage{gensymb}

\usepackage{undertilde}
\makeatletter
\def\simgt{\mathrel{\lower2.5pt\vbox{\lineskip=0pt\baselineskip=0pt
           \hbox{$>$}\hbox{$\sim$}}}}
\def\simlt{\mathrel{\lower2.5pt\vbox{\lineskip=0pt\baselineskip=0pt
           \hbox{$<$}\hbox{$\sim$}}}}
\makeatother

\newcommand{\be}{\begin{equation}}
\newcommand{\ee}{\end{equation}}
\newcommand{\bea}{\begin{eqnarray}}
\newcommand{\eea}{\end{eqnarray}}

\begin{document}

\title{Multi-Step Cascade Annihilations of Dark Matter and the Galactic Center Excess}

\author{Gilly Elor}
\author{Nicholas L. Rodd}
\author{Tracy R. Slatyer}
\affiliation{Center for Theoretical Physics, Massachusetts Institute of Technology, Cambridge, MA}

\begin{abstract} If dark matter is embedded in a non-trivial dark sector, it may annihilate and decay to lighter dark-sector states which subsequently decay to the Standard Model. Such scenarios -- with annihilation followed by cascading dark-sector decays -- can explain the apparent excess GeV gamma-rays identified in the central Milky Way, while evading bounds from dark matter direct detection experiments. Each `step' in the cascade will modify the observable signatures of dark matter annihilation and decay, shifting the resulting photons and other final state particles to lower energies and broadening their spectra. We explore, in a model-independent way, the effect of multi-step dark-sector cascades on the preferred regions of parameter space to explain the GeV excess. We find that the broadening effects of multi-step cascades can admit final states dominated by particles that would usually produce too sharply peaked photon spectra; in general, if the cascades are hierarchical (each particle decays to substantially lighter particles), the preferred mass range for the dark matter is in all cases 20-150 GeV. Decay chains that have nearly-degenerate steps, where the products are close to half the mass of the progenitor, can admit much higher DM masses. We map out the region of mass/cross-section parameter space where cascades (degenerate, hierarchical or a combination) can fit the signal, for a range of final states. In the current work, we study multi-step cascades in the context of explaining the GeV excess, but many aspects of our results are general and can be extended to other applications.
\end{abstract}

%12.60.Jv, 11.30.Fs, 11.30.Qc}

\maketitle

\section{Introduction}

Over the past five years, numerous independent studies have confirmed a flux of few-GeV gamma rays from the inner Milky Way, steeply peaked toward the Galactic Center, that is not captured by models for the known diffuse backgrounds \cite{Goodenough:2009gk,Hooper:2010mq,Boyarsky:2010dr,Hooper:2011ti,Abazajian:2012pn,Hooper:2013rwa,Gordon:2013vta,Huang:2013pda,Abazajian:2014fta,2014arXiv1402.6703D,Calore:2014xka}. This ``Galactic Center excess'' (GCE), detected using public data from the {\it Fermi} Gamma-Ray Space Telescope, has a spatial morphology well described by the square of a generalized Navarro-Frenk-White (NFW) profile, projected along the line of sight. Furthermore, it is highly spherically symmetric,  centered on the Galactic Center (GC), and extends at least 10 degrees from the GC \cite{2014arXiv1402.6703D};\footnote{This analysis exploited improvements to the \emph{Fermi} point spread function as described in \cite{2014arXiv1406.0507P}.} these conclusions remain unchanged when accounting for systematic uncertainties in the modeling of the diffuse backgrounds \cite{Calore:2014xka}. These spatial properties suggest the excess emission could arise from the annihilation of dark matter (DM) with an NFW-like density profile. Competing interpretations include a transient event at the GC producing high-energy cosmic rays that subsequently yield few-GeV gamma rays by scattering processes \cite{Petrovic:2014uda, Carlson:2014cwa}, or a population of many unresolved millisecond pulsars (MSPs) (e.g. \cite{Abazajian:2010zy, Gordon:2013vta}). However, these interpretations face significant challenges: it is unclear whether the proposed outflow models can match the spectrum and morphology of the excess \cite{LINDENTALK} (see also \cite{Macias:2013vya,Gordon:2014gya}), and estimates of the MSP population in the region of interest consistently underpredict the signal by an order of magnitude \cite{Hooper:2013nhl, Cholis:2014lta}.

Models where DM annihilates with a roughly thermal cross-section and has a mass of order several tens of GeV can readily account for the spectrum and size of the excess. However, when embedded in even a simplified DM model, there are often powerful constraints on these scenarios from direct detection and collider bounds (e.g. \cite{Alves:2014yha, Berlin:2014tja}). While UV-complete models where the DM annihilates directly to Standard Model (SM) particles do exist (e.g. \cite{Ipek:2014gua, Cheung:2014lqa, Gherghetta:2015ysa}), the constraints are much more easily evaded if the DM produces gamma-rays via a cascade process \cite{Pospelov:2007mp, Martin:2014sxa,Abdullah:2014lla,Ko:2014gha,Freytsis:2014sua}. In such scenarios, the DM is secluded in its own hidden dark sector, and first annihilates to other dark sector particles; these mediators subsequently decay into SM particles that produce gamma-rays.\footnote{Annihilation into the dark sector can also lead to a novel spatial distribution for the signal \cite{Rothstein:2009pm}, but the GCE favors a cuspy morphology, so in this work we assume all decays are prompt.}

The presence of an intermediate step between DM annihilation and the production of SM particles broadens the spectrum of SM particles produced, and consequently also broadens the resulting gamma-ray spectrum, unless the mediator is degenerate in mass with either the DM or  the total mass of the SM decay products. The gamma-ray multiplicity is increased by a factor of two, if each mediator decays into two SM particles, and the typical energy of the gamma-rays is reduced accordingly. Thus cascade models for the excess generically tend to accommodate:

\begin{itemize}
\item Higher DM masses,
\item Decays of the mediator to SM final states whose decays produce a more sharply peaked gamma-ray spectrum than favored by direct annihilation.
\end{itemize} 

In general, there may be more than one decay step within the dark sector; the dominant annihilation of the DM need not be to the lightest dark sector particle (e.g. \cite{Baumgart:2009tn, Nomura:2008ru}). If couplings within the dark sector are stronger than couplings between the sectors, dark sector particles will preferentially decay within the dark sector, with decays to the SM only occurring when no other states are available. Regardless of the model under consideration, in the absence of a mass degeneracy, each decay will increase the final gamma-ray multiplicity, decrease the typical gamma-ray energy, and broaden the spectrum (in the presence of a mass degeneracy only the first two effects will occur). Accordingly, long decay chains could potentially permit much heavier DM to explain the GCE, or favor decays to different SM states. In a sense, this description also characterizes the known decays of SM particles; final states whose decays produce gamma-rays through a lengthy cascade will generate a broader spectrum with a lower-energy peak, compared to final states that generate gamma-rays via a short cascade (we discuss this further in Sec. \ref{sec:results}).

It is this possibility of multi-step dark sector cascades that we explore in this work. For simplicity, we consider the case where all dark-sector particles involved in the cascade (except possibly the DM itself) are scalars - we briefly discuss the case of non-scalar mediators in Sec.~\ref{sec:generalcascade}. In this case, the results are largely independent of the details of the dark sector. The DM pair-annihilates into two scalar mediators which subsequently undergo a multi-step cascade in the dark sector, eventually producing a dark-sector state (with high multiplicity) that decays to the SM:
\be\begin{aligned}
\chi \chi \rightarrow \phi_n \phi_n &\rightarrow 2 \times \phi_{n-1} \phi_{n-1} \rightarrow . . . \\
 &\rightarrow 2^{n-1} \times \phi_1 \phi_1 \rightarrow 2^{n} \times f \bar{f}\,.
\label{eq:cascade}
\end{aligned}\ee
Here $f \bar{f}$ are SM lepton or quark pairs, which can subsequently decay; the decays shown above may also produce photons in the final step via final state radiation (FSR). By fitting the resulting photon spectrum to the GCE, we determine the allowed values of cross-section and DM mass for cascades with one to six steps, for a variety of SM final states. Provided that the masses of the particles at each step in the cascade are not near-degenerate, the final spectrum of gamma-rays becomes nearly independent of the exact masses at each step. This assumption is not limiting, as results for the quality of fit for the more general case of non-hierarchical cascades (with nearly-degenerate steps) can be simply extracted from results derived assuming a large hierarchy.

In Sec.~\ref{sec:methods} we outline the determination of the photon spectrum for an $n$-step cascade with specified SM final state, and discuss the procedure used to compare such a spectrum to the GCE. We present sample results of these fits in Sec.~\ref{sec:results} under certain assumptions. Section~\ref{sec:generalcascade} extends our results for general cascades, and contains our complete fit results. In Sec.~\ref{sec:signalsconstraints} we outline the existing experimental constraints a complete  model for the GCE via cascade decays would need to satisfy. We present our conclusions in Sec.~\ref{sec:conclusion}. In the appendices we provide additional details of our methodology and discuss some further model-dependent considerations. 

\section{Methodology}
\label{sec:methods}
The photon flux generated by the annihilations of self-conjugate DM\footnote{As discussed in Appendix~\ref{app:zerostep}, our results can be readily translated to the case of decays, although the steeply peaked morphology of the GCE disfavors this interpretation.} as a function of the direction observed in the sky, is given by:
\be
\Phi\left(E_\gamma, l, b \right) = \frac{\langle \sigma v \rangle}{8 \pi m_{\chi}^2} \frac{dN_\gamma}{dE_\gamma} J\left(l,b\right)\,,
\label{eq:flux}
\ee
where $\langle \sigma v \rangle$ is the thermally averaged annihilation cross-section, $m_\chi$ is the DM mass, and $dN_\gamma / dE_\gamma$ is the photon spectrum per DM annihilation, which has contributions from FSR and from the decay of the leptons or quarks and their subsequent hadronization products. The $J$-factor, the integral of DM density squared along the line-of-sight, is a function of the observed direction in the sky expressed in terms of Galactic coordinates $l$ and $b$:
\be
J\left(l,b\right) =   \int_{0}^{\infty}  \rho^2\left(\sqrt{s^2 - 2 r_{\odot} s  \cos l \cos b + r_{\odot}^2}  \right) ds\,,
\label{eq:Jfactor}
\ee
where $r_{\odot}\approx8.5$ kpc is the distance from the Sun to the Galactic Center, and  $s$ parametrizes the integral along the line-of-sight. We parameterize the DM density by a generalized NFW halo profile \cite{Navarro:1995iw,Navarro:1996gj}:
\be
\rho \left( r, \gamma \right) = \rho_0 \frac{(r/r_s)^{-\gamma}}{\left(1 + r/r_s \right)^{3-\gamma}}\,.
\ee
Here we use $r_s = 20$ kpc, $\rho_0 = 0.4$ GeV/cm$^3$ and $\gamma = 1.2$, following \cite{Calore:2014xka}, as we will compare our models to the data using the spectrum and covariance matrix determined by that work.

We focus on $n$-step cascades ending in $\phi_1 \rightarrow f \bar{f}$, where $f \bar{f}$ is a pair of electrons, muons, taus or $b$-quarks. Other SM final states are possible, of course, but these cases span the range from steeply peaked photon spectra close to the DM mass through to the lower-energy and broader spectra characteristic of annihilation to hadrons. In order to generate the cascade spectrum, we first start with the result from direct DM annihilation, which is equivalent to the spectrum from $\phi_1$ decay (in the $\phi_1$ rest frame) if the DM mass is half the $\phi_1$ mass. For the case of electrons or muons we determine this spectrum analytically using the results of \cite{Mardon:2009rc}, whilst for taus and $b$-quarks the results are simulated in \texttt{Pythia8} \cite{Sjostrand:2007gs}. We have relegated the details of calculating these spectra to Appendix~\ref{app:zerostep}.

We denote the spectrum obtained at this ``0th step'' by $dN_\gamma / dx_{0}$, where $x_0 = 2E_0 / m_1$, $m_1$ is the mass of $\phi_1$ and $E_0$ is the energy of the photon in the $\phi_1$ rest frame. The shape of the photon spectrum is determined by the identity of the final state particle $f$ and also the ratio $\epsilon_f = 2m_f / m_1$. In the limit where the decay of $\phi_1$ is dominated by a two-body final state (at least for the purposes of photon production), the photon spectrum converges to a constant shape (as a function of $x_0$) as $\epsilon_f \rightarrow 0$ and the $f \bar{f}$ become highly relativistic. However, final state radiation (FSR) and hadronization depend on the energy of the $f \bar{f}$ products of the $\phi_1$ decay in the $\phi_1$ rest frame, so in cases where these effects dominate, the dependence of the photon spectrum on $\epsilon_f$ is more complex.

\begin{figure}[t!]
\centering
\includegraphics[scale=0.7]{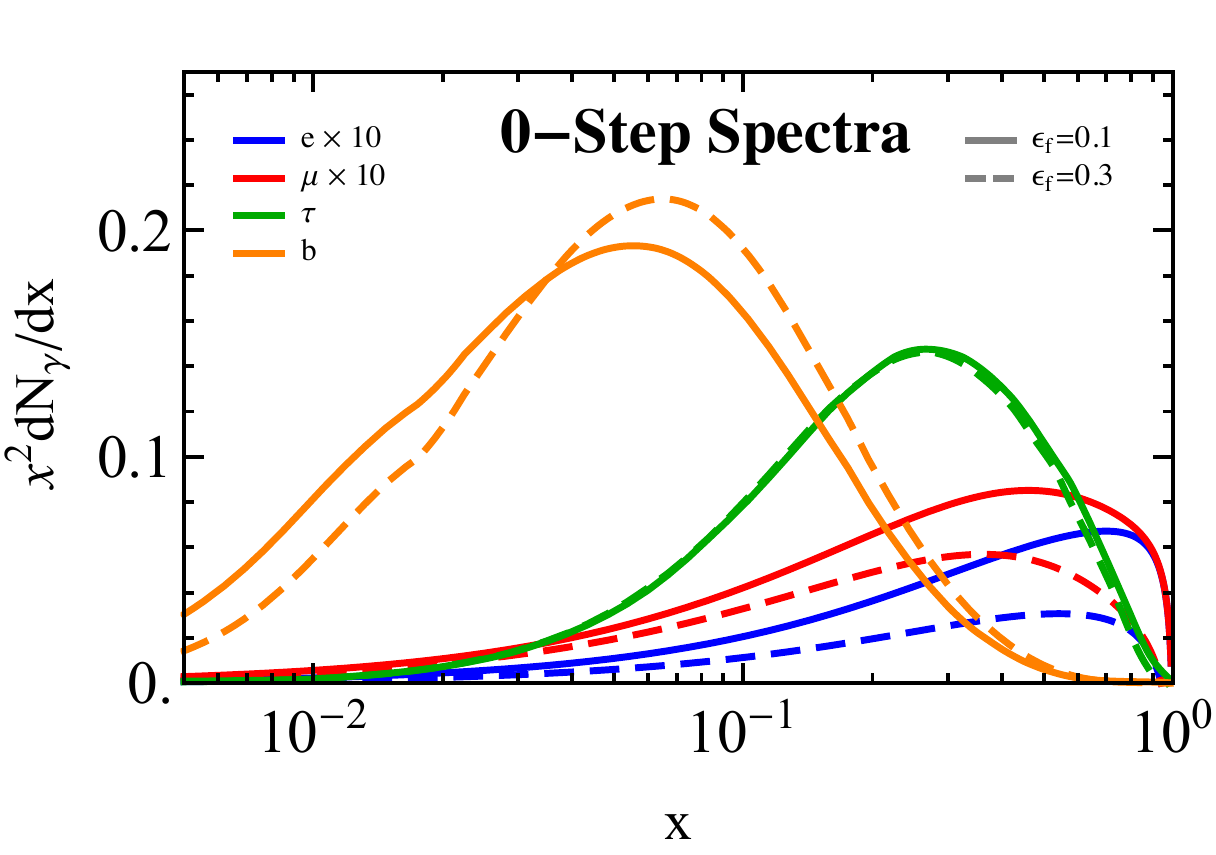}
\caption{\footnotesize{0th step (direct annihilation) photon spectra $dN_\gamma/dx_{0}$ for $\phi_1$ decaying to $(e, \mu, \tau, b)$ in (blue, red, green, orange). Solid curves correspond to $\epsilon_f=0.1$, and dashed to $\epsilon_f=0.3$. The electron and muon spectra have been magnified by a factor of ten to appear comparable to the taus and $b$s.}}
\label{fig:0step}
\end{figure}

In Fig.~\ref{fig:0step} we show $dN_\gamma/dx_{0}$ per annihilation for the four different final states we considered, for $\epsilon_f = 0.1$ and $\epsilon_f = 0.3$. The photon spectra from electron and muon production are dominated by FSR, whereas for $b$-quarks fragmentation and hadronization are important. In the photon spectrum from taus, these effects are subdominant and so the impact of varying $\epsilon_f$ is minimal. Note that the spectrum for $b$-quarks is peaked at a significantly lower $x$, highlighting why models with this final state tend to accommodate higher DM masses.

Given the 0-step spectrum, determining the photon spectrum from an $n$-step cascade is particularly simple in the case of scalar mediators,\footnote{We discuss the case of vector mediators in Sec.~\ref{sec:generalcascade}.} where the calculation essentially reduces to Lorentz-boosting the photon spectrum up the ladder of particles appearing in the cascade. We review this calculation in Appendix~\ref{app:boost}. As observed in \cite{Mardon:2009rc}, in the case of large mass hierarchies between the steps in the cascade, the final photon spectrum can be simplified even further, as we now outline. 

Consider the $i$th step in the cascade, where the decay is $\phi_{i+1} \to \phi_i \phi_i$. Let us define $\epsilon_i = 2m_i/m_{i+1}$, and assume $\epsilon_i \ll 1$.\footnote{Note that the earlier-defined $\epsilon_f$ parameter does \emph{not} function in exactly the same way as these $\epsilon_i$ parameters: $\epsilon_f$ fully parameterizes the photon spectrum associated with production and decay of the SM particles, whereas the $\epsilon_i$ only describe Lorentz boosts.}  Suppose the photon spectrum from decay of a single $\phi_i$ (and the subsequent cascade), in the rest frame of the $\phi_i$ particle, is known and denoted by $dN_\gamma/dx_{i-1}$. Then, in the presence of a large mass hierarchy, the decay of $\phi_{i+1}$ produces two highly relativistic $\phi_i$ particles, each (in the rest frame of the $\phi_{i+1}$) carrying energy equal to $m_{i+1}/2 = m_i/\epsilon_i$. The photon spectrum in the rest frame of the $\phi_{i+1}$ is then given by a Lorentz boost (see Appendix~\ref{app:boost}), and in the limit $\epsilon_i \ll 1$ takes the simple form \cite{Mardon:2009rc}:
\be
\frac{dN_\gamma}{dx_i} = 2 \int_{x_i}^{1} \frac{dx_{i-1}}{x_{i-1}} \frac{dN_\gamma}{dx_{i-1}} + \mathcal{O}(\epsilon_i^2)\,.
\label{eq:boosteq}
\ee
Here we have introduced the dimensionless variable $x_i = 2 E_i / m_{i+1}$, where $E_i$ is the photon energy in the $\phi_{i+1}$ rest frame. Following this, once we know the 0-step spectrum we can iteratively derive the $n$-step result. The error introduced by this assumption is $\mathcal{O}(\epsilon_i^2)$, as we quantify in Appendix~\ref{app:boost}. 

Beyond simplifying calculations, the large hierarchy approximation is also convenient for the following two reasons. Firstly in this limit, we can specify the shape of the spectrum simply by the identity of the final state $f$, the value of $\epsilon_f$, and finally the number of steps $n$. This is in contrast to the many possible parameters that could be present in a generic cascade. Secondly, as we will elaborate further in Sec.~\ref{sec:generalcascade}, it is also possible to read off the results for a generic hierarchy once we know the small $\epsilon_i$ result, making the assumption less limiting than it would initially appear. In particular in the limit when the masses become degenerate ($\epsilon_i \to 1$), the $\phi_i$'s are produced at rest. When they subsequently decay, there is no boost to the $\phi_{i+1}$ rest frame, and so an $n$-step cascade effectively reduces to a hierarchical $(n-1)$-step cascade, except for the additional final state multiplicity.  
 
\begin{figure}[t!]
\centering
\includegraphics[scale=0.7]{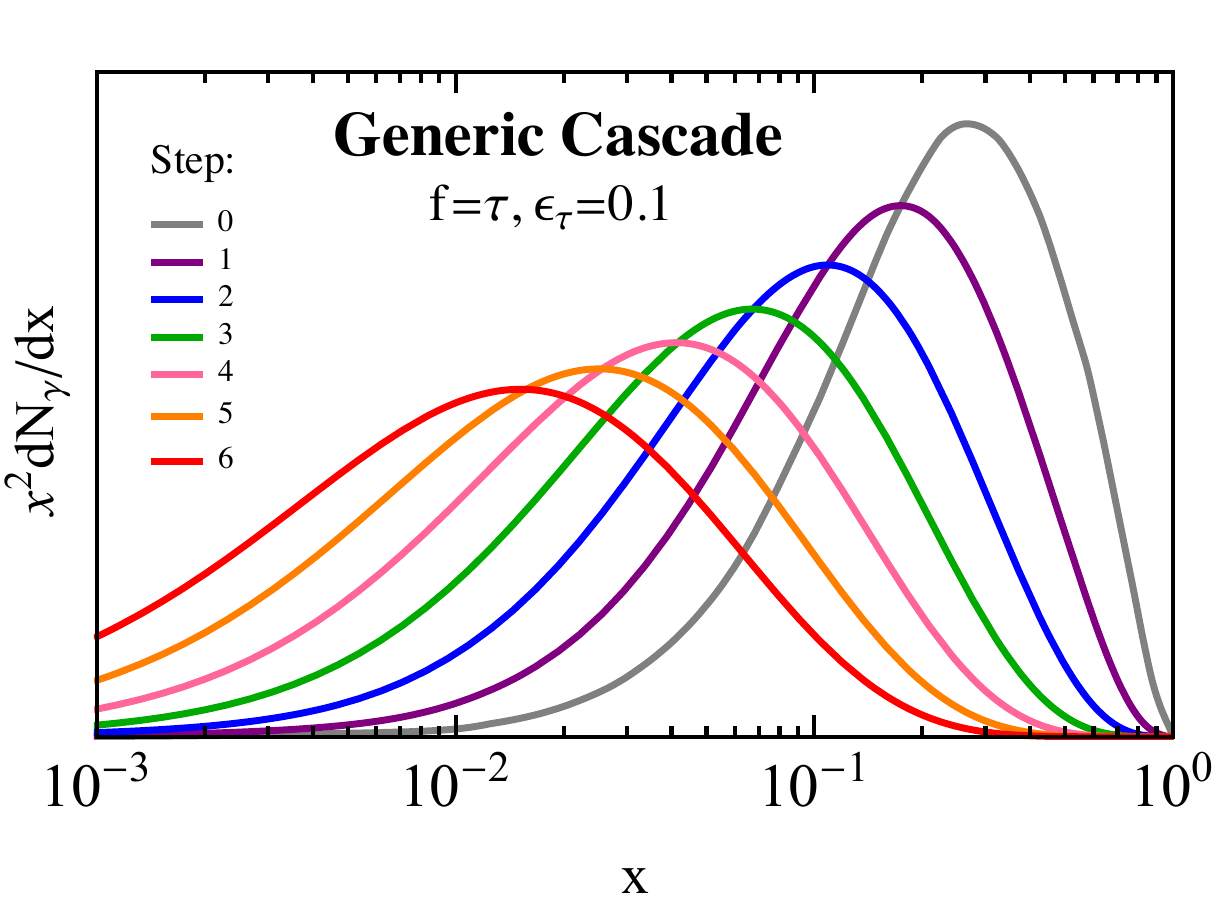}
\caption{\footnotesize{An example photon spectrum from direct annihilation to taus (grey) and hierarchical cascades with $n$ = (1,2,3,4,5,6) steps, corresponding to (purple, blue, green, pink, orange, red) curves. The presence of each additional step in the cascade acts to broaden and soften the spectrum, and shift the peak to lower masses. All spectra are per annihilation.}}
\label{fig:GeneralSpectrum}
\end{figure}

The Galactic frame is approximately the rest frame of the (cold) DM; consequently, to determine the measured photon spectrum, we need to calculate the photon spectrum in the rest frame of the original DM particles. For an $n$-step cascade, this will involve $n$ such convolutions, starting from the $dN_\gamma/dx_0$ 0-step spectrum, where the highest mass scale in the cascade will be $m_{i=n} = 2 m_\chi$. Thus $x_{i=n} = E_n / m_{\chi}$, and the Galactic-frame photon spectrum will be $dN_\gamma/dx_n = m_{\chi} dN_\gamma / dE_n$. Fig.~\ref{fig:GeneralSpectrum} shows the resulting spectrum for a 0-6 step cascade in the case of final state taus with $\epsilon_{\tau}=0.1$. Each step in the cascade broadens out and softens the spectrum, and similar behaviour is seen for other final states.

In order to determine the favored parameter space, for a given choice of $f$, $\epsilon_f$, and number of steps in the cascade $n$, we vary $m_\chi$ and an overall normalization parameter $\eta$ (proportional to $\langle \sigma v \rangle /m_\chi^2$, as we will see below) and compare the model to the data using the spectrum and covariance matrix of \cite{Calore:2014xka}. In detail we calculate $\chi^2$ according to:
\be
\chi^2 = \sum_{ij} \left(\mathcal{N}_{i,\textrm{model}} - \mathcal{N}_{i,\textrm{data}} \right) C^{-1}_{ij} \left(\mathcal{N}_{j,\textrm{model}} - \mathcal{N}_{j,\textrm{data}} \right)\,,
\label{eq:chi}
\ee
where
\bea
&\mathcal{N}_{i,\rm{model}}& =  \left( \frac{\eta}{m_{\chi}}E_n^2\frac{dN}{dx_n}\right)_{i,\rm{model}} \\
&\mathcal{N}_{i,\rm{data}}& = \left(E^2 \frac{dN}{dE}\right)_{i,\rm{data}}
\label{eq:chi2}
\eea
and both model and data are expressed in units of GeV/cm$^2$/s/sr averaged over the region of interest. Here the $C^{-1}_{ij}$ are elements of the inverse covariance matrix, which together with the data points are taken from \cite{Calore:2014xka}. By Eq.~\ref{eq:flux}, the fitted normalization $\eta$ is related to the DM mass and the J-factor by:
\be
\langle \sigma v \rangle = \frac{8 \pi m_{\chi}^2 \eta} {J_{\rm norm}}\,.
\label{eq:xsecnorm}
\ee
For consistency with the spectrum normalization of \cite{Calore:2014xka} the J-factor is averaged over the ROI $|l| \leq 20^\circ$ and $2^\circ \leq |b| \leq 20^\circ$, so that:
\be\begin{aligned}
J_{\rm norm} &=  \int_{\rm ROI} d \Omega J\left(l,b\right)  / \int_{\rm ROI} d\Omega \\
&\sim 2.0618 \times 10^{23}~{\rm GeV}^2{\rm cm}^{-5}.
\label{eq:Jnorm}
\end{aligned}\ee 
(Note that $d\Omega = dl d\sin b$, not $dl d\cos b$, since $b$ measures the angle from the Galactic equator, not the north pole.)

\emph{Self-Consistency Requirements:} The procedure outlined above treats $m_\chi$ as a free parameter that can be adjusted to modify the 0-step spectrum; the fit only uses the shape of the spectrum provided by the 0-step result and the boost of Eq.~\ref{eq:boosteq}. However, there is an additional condition required for a cascade scenario to be physically self-consistent: the mass hierarchy between the DM mass and the particles produced in the final state must be sufficiently large to accommodate the specified number of steps. Equivalently, there is a hard upper limit on the number of steps allowed, for a given DM mass and final state.

Recall that for an $n$-step cascade ending in a final state $f$, we defined $\epsilon_f = 2 m_f/m_1$, $\epsilon_1 = 2m_1/m_2$, $\epsilon_2 = 2m_2/m_3$ all the way up to $\epsilon_n=m_n/m_{\chi}$. Combining these, the DM mass is given in terms of $m_f$ and the $\epsilon$ factors by:
\be
m_\chi = 2^n \frac{m_f}{\epsilon_f  \epsilon_1  \epsilon_2  . . .  \epsilon_n}\,,
\label{eq:mDM}
\ee
If the $\epsilon_i$ factors are allowed to float, we can still say that $0 < \epsilon_i \leq 1$ in all cases (since each decaying particle must have enough mass to provide the rest masses of the decay products), setting a strict lower bound on the DM mass of:
\be m_\chi \geq 2^n m_f/\epsilon_f\,. \label{eq:kinematic} \ee 
In the remainder of this article we refer to this bound as a ``self-consistency'' condition or defining ``kinematically allowed'' masses. For consistency with the assumption of hierarchical decays (i.e. $\epsilon_i \ll 1$), the true bound on $m_\chi$ will in general be somewhat stronger than this conservative estimate (although as we will discuss in Sec.~\ref{sec:generalcascade}, $\epsilon_i$ can become quite close to 1 before significantly modifying the fit relative to the $\epsilon_i \rightarrow 0$ case). 

\section{Results With the Assumption of Large Hierarchies}
\label{sec:results}
\begin{figure*}[t!]
\centering
\begin{minipage}{.45\textwidth}
  \centering
  \includegraphics[scale=0.65]{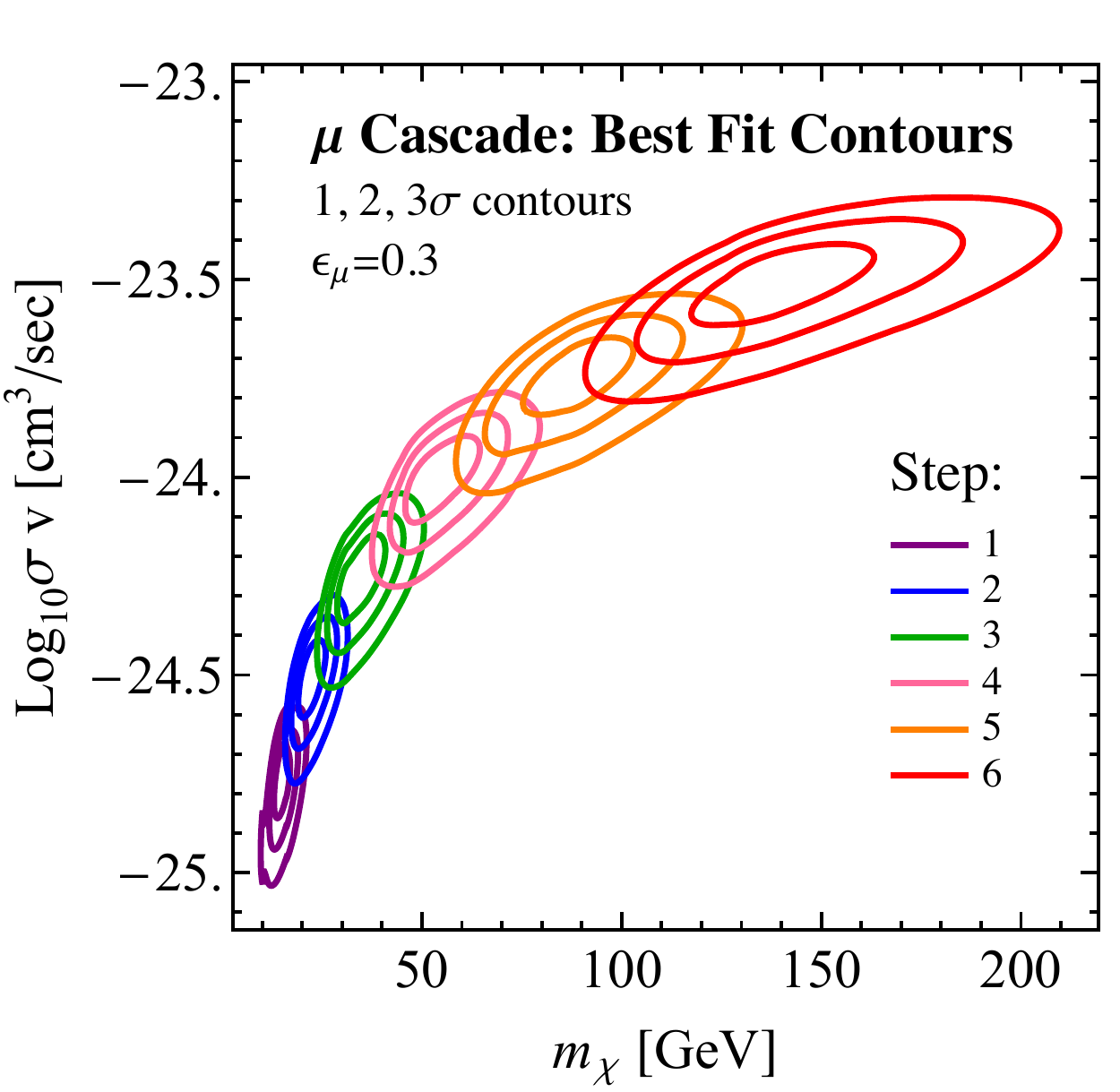}
  \captionof{figure}{\footnotesize{Contours of $\Delta \chi^2$ from the best-fit point (for a given step number $n$) corresponding to 1, 2 and 3$\sigma$ for final state $\mu$'s, with $\epsilon_{\mu} = 0.3$. The purple, blue, green, pink, orange and red colors correspond to $n =$ 1, 2, 3, 4, 5 and 6 steps in the cascades to final state $\mu$'s. Here we have fixed $\epsilon_{\mu} = 0.3$ and fit over the range 0.5 GeV $\leq E_\gamma \leq$ 300 GeV.}}
  \label{fig:TauChiPlot}
\end{minipage}
\hspace{0.4in}
\begin{minipage}{.45\textwidth}
\vspace{0.05in}
  \centering
  \includegraphics[scale=0.63]{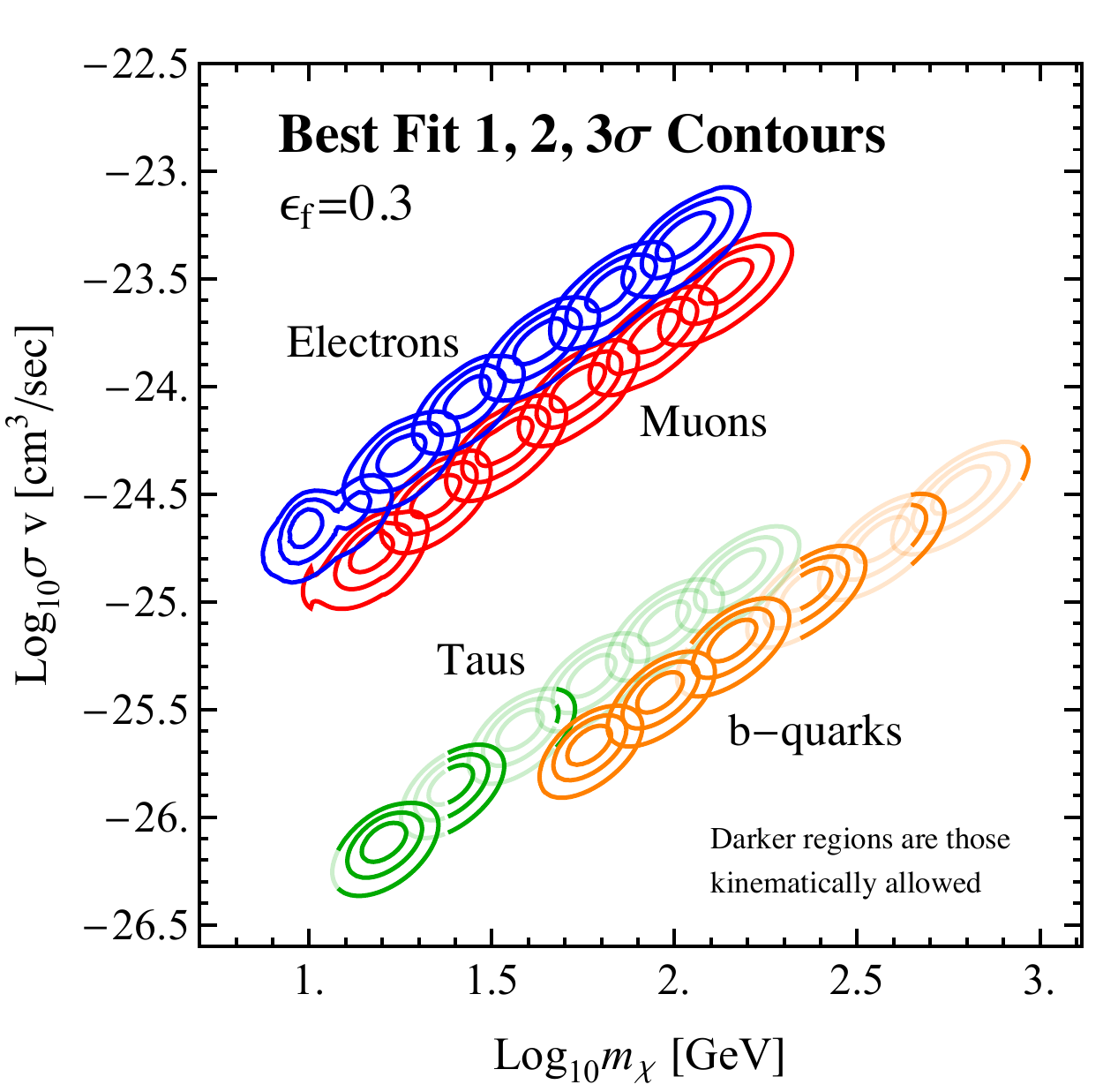}
  \captionof{figure}{\footnotesize{Contours of $\Delta \chi^2$ corresponding to 1, 2 and 3$\sigma$ for $n=1-6$ steps for $e$, $\mu$, $\tau$ and $b$ final states with $\epsilon_f=0.3$. The fit is performed over the range 0.5 GeV $\leq E_\gamma \leq$ 300 GeV. The best fit point of each step for all four final states follows a power law relation between $m_\chi$ and $\langle \sigma v \rangle$, with index $\sim 1.3$. Only the darker regions are kinematically allowed. See text for details.}}
  \label{fig:AllStatesLogChiPlot}
\end{minipage}
\end{figure*}

Here we present the results from the fits performed using the procedure outlined in the previous section. Assuming hierarchical cascades, we perform fits for four different final states -- electrons, muons, taus, and $b$-quarks -- and fit over the photon energy range $0.5~\textrm{GeV} \leq E_{\gamma} \leq 300~\textrm{GeV}$.\footnote{By default, we omit the low energy data points with $0.3~\textrm{GeV} \leq E_{\gamma} \leq 0.5~\textrm{GeV}$, as in this region the spectrum suffers larger uncertainties under variations of the background modeling, and the preferred value of the NFW $\gamma$ parameter is not robust \cite{2014arXiv1402.6703D}. We have confirmed that including these low-energy data points has little impact on our results.} Later in this section we discuss the effects of cutting out high energy data points, and how the fits would change if we only considered statistical uncertainties. 

In Fig.~\ref{fig:TauChiPlot} we show a sample result, in which we plot $\Delta \chi^2$ 1, 2 and 3$\sigma$ contours in $\left(m_{\chi}, \langle \sigma v \rangle \right)$ space for 1-6 step cascades ending in muons with $\epsilon_{\mu} = 0.3$. The trend in the best fit point for each step is as expected. Recall the generic behavior illustrated in Fig.~\ref{fig:GeneralSpectrum}; each progressive step in the cascade acts to reduce the height of the peak and shift it to lower masses. Therefore higher steps in the cascades will be better fit by larger DM mass and cross-section as is indeed the case in Fig.~\ref{fig:TauChiPlot}. The larger cross-section results from an interplay of effects as can be seen from Eq.~\ref{eq:xsecnorm}: an increased DM mass leads to a lower number density and hence a higher cross-section (scaling as $m_\chi^2$), but the increased power per annihilation implies a lower $\eta$ (adding a factor of $m_\chi^{-1}$), and finally the reduced height of the peak in the dimensionless spectrum for higher steps (as shown in Fig.~\ref{fig:GeneralSpectrum}) requires a larger $\eta$.

\begin{figure*}[t!]
\centering
\begin{tabular}{c}
\includegraphics[scale=0.57]{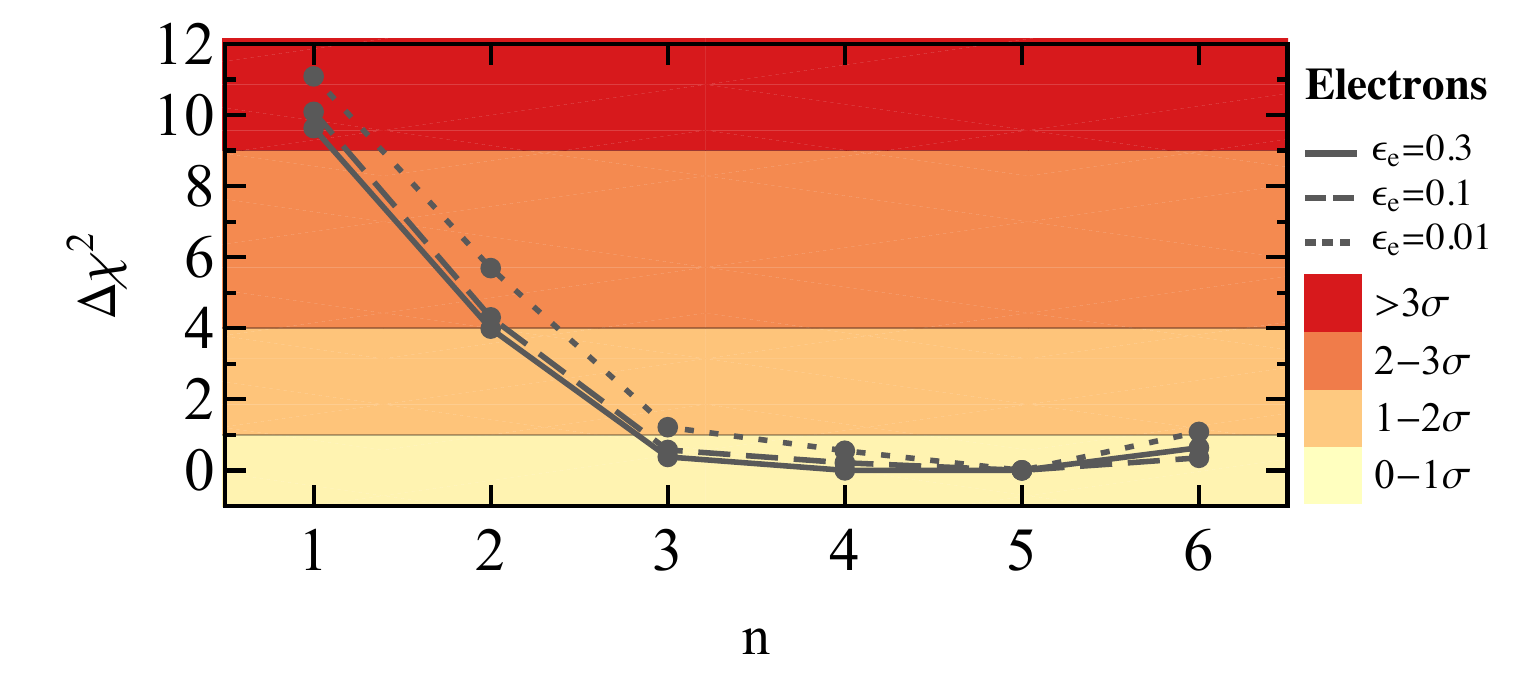} \hspace{0.1in}
\includegraphics[scale=0.57]{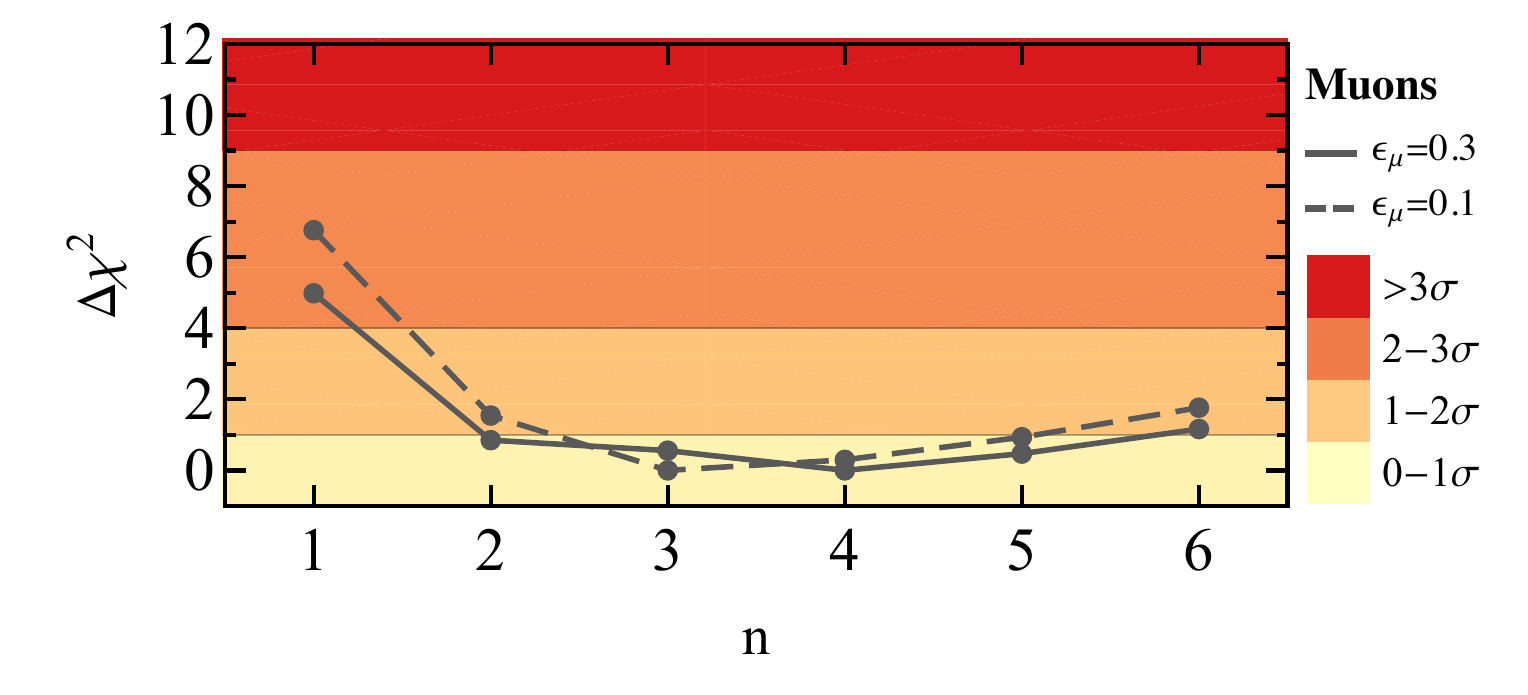}\\
\includegraphics[scale=0.57]{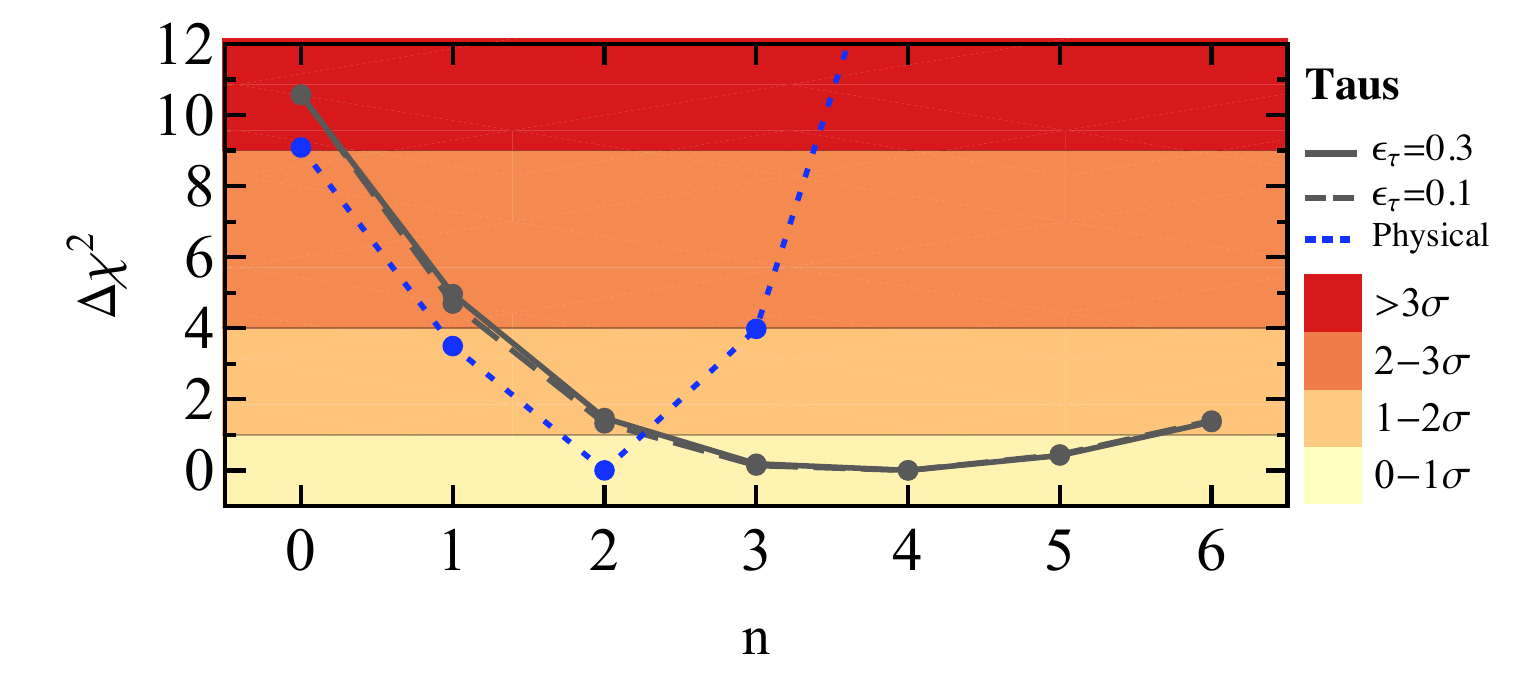} \hspace{0.1in}
\includegraphics[scale=0.57]{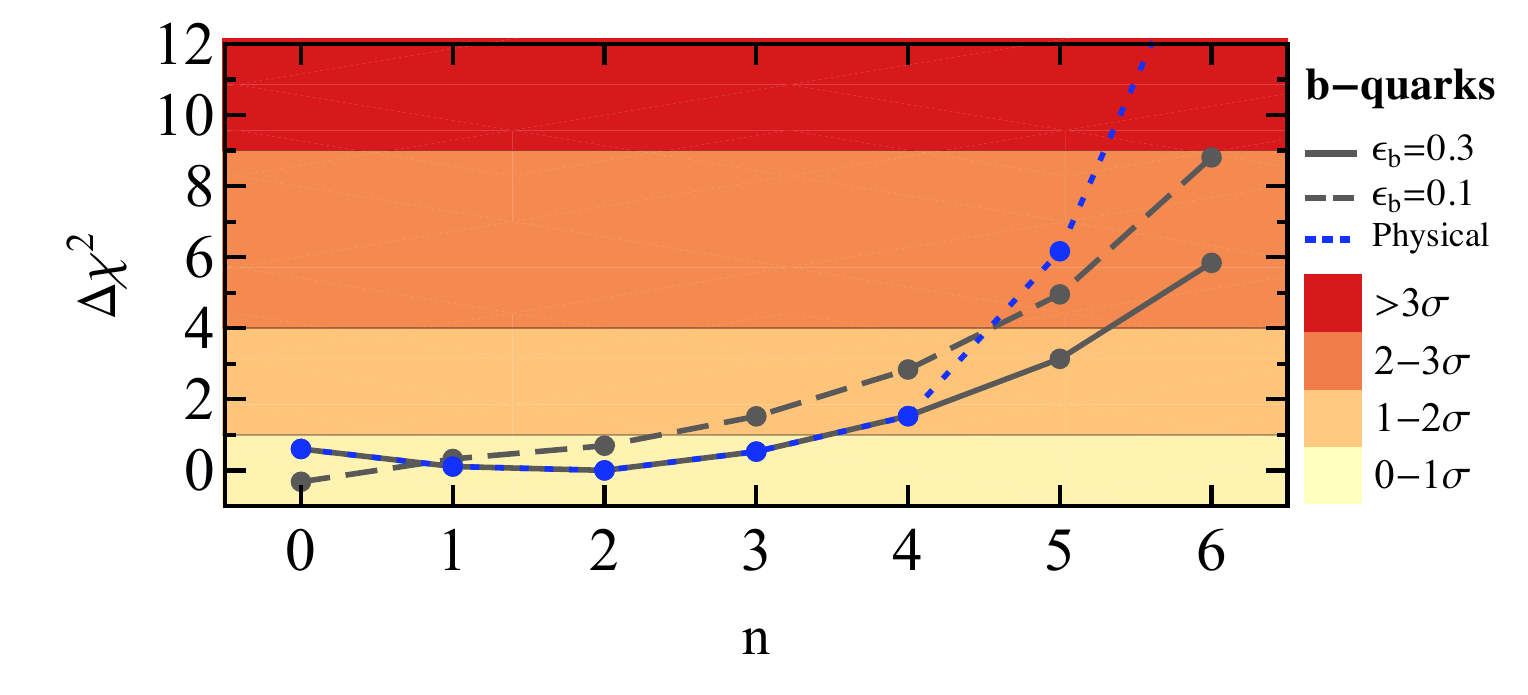}
\end{tabular}
\caption{\footnotesize{Clockwise panels show the overall best fit for DM annihilating through an $n$-step cascade to electron, muon, $b$-quark and tau final states. The grey solid, dashed (and dotted) lines correspond to the $\Delta \chi^2$ between the best fit at that step, and the best fit for all $n$, for $\epsilon_{f} = 0.3,0.1$ (and $0.01$) respectively. In the case of tau and $b$-quark final states, the blue dotted curves, denoted `physical,' correspond to the case where only kinematically allowed  (self-consistent) masses are considered as per the discussion in Sec.~\ref{sec:methods} (we set $\epsilon_f = 0.3$ for these curves). Note that in the case of taus, the ``physical'' best-fit points for 0 and 1 steps have the same $\chi^2$ as the best-fit points when ``unphysical'' scenarios are allowed, but as the overall best fit is different (with higher $\chi^2$) their $\Delta \chi^2$ is lower. The shaded bands correspond to the quality of fit. 0-step results are not included for electrons and muons, as these fits are poor and have $\Delta \chi^2$ values well above the plotted $y$-axis. Electrons, muons and taus prefer longer 3-5 step cascades, whilst annihilations to $b$-quarks prefer shorter 0-2 step cascades. This is not surprising, since as has been already pointed out in the literature, $b$-quark final states are preferred for direct annihilations. Non-integer values of $n$ can be associated with cascades containing steps with one or more large $\epsilon_i$, as discussed in Sec.~\ref{sec:generalcascade}.}}
\label{fig:BestFit}
\end{figure*}

\begin{figure}[t!]
\centering
\includegraphics[scale=0.65]{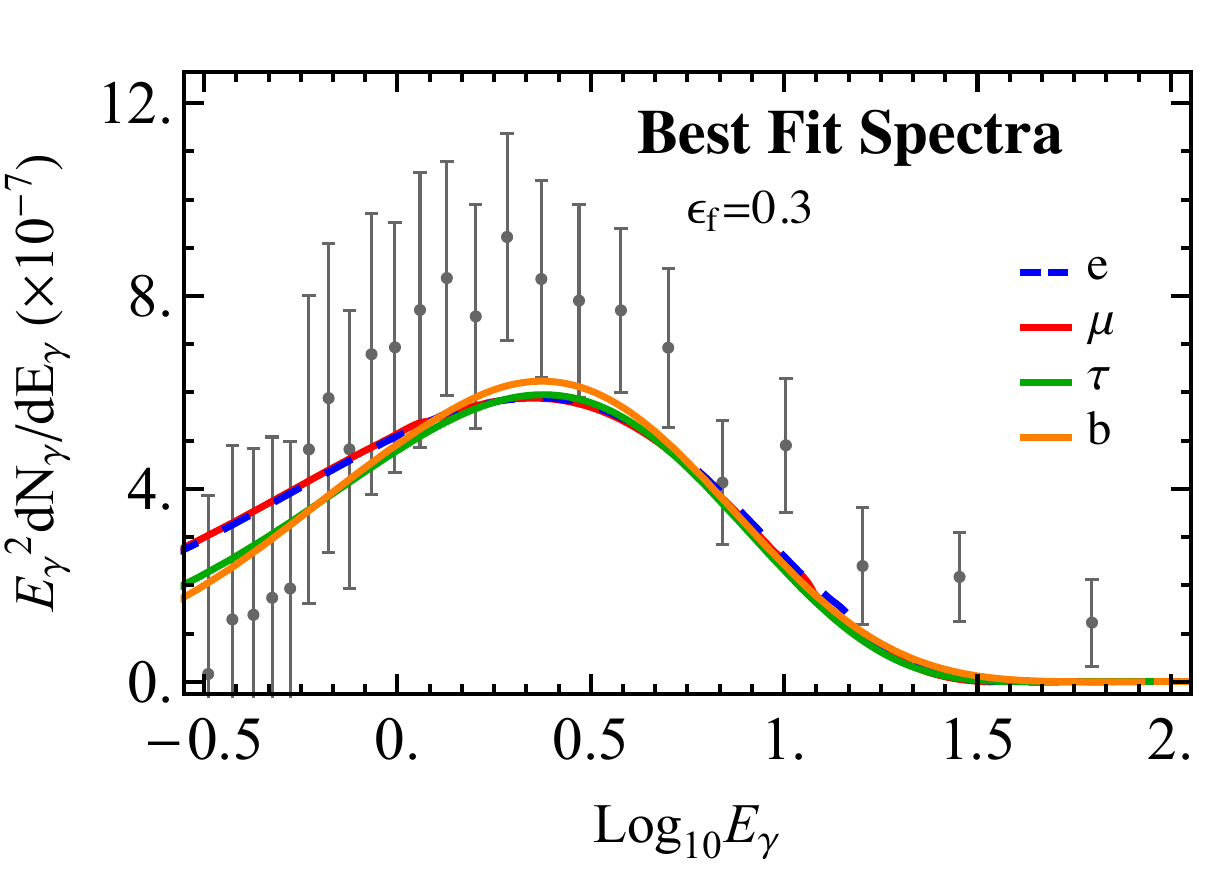}
\caption{\footnotesize{The blue, red, green and orange curves correspond to the overall best fit spectrum for e, $\mu$, $\tau$ and $b$-quarks as determined from Fig.~\ref{fig:BestFit}. Overlaid are the data points and systematic errors from \cite{Calore:2014xka}. Note that due to correlations between energies, the best fit curves are not what would be naively expected if only statistical errors were present.}}
\label{fig:Spectrum0p3Full}
\end{figure}

In Fig.~\ref{fig:AllStatesLogChiPlot} we show the corresponding $\Delta \chi^2$ contours for electron, muon, tau, and $b$-quark final states, again fixing $\epsilon_f=0.3$. The best-fit mass and cross-section for each of the final states are empirically found to follow an approximate power law with $\langle \sigma v \rangle \propto m_\chi^{1.3}$. As discussed above we would expect $\langle \sigma v \rangle \propto m_\chi$ if the spectrum did not change in shape (simply being rescaled proportionally to $m_\chi$ to ensure energy conservation); the additional $m_\chi^{0.3}$ scaling factor reflects the change in shape of the spectrum.

As discussed above, for a given DM mass and final-state fermion with mass $m_f$, there is an absolute upper limit on the number of steps allowed in a cascade, since every step corresponds to a change in mass scale of at least a factor of 2. In Fig. \ref{fig:AllStatesLogChiPlot}, we show the contours if the limitation of Eq.~\ref{eq:kinematic} is \emph{ignored}, since this conveys information on the mass scale and number of steps at which the broadness of the spectrum best matches the data; however, the mass values that violate this condition and so do not represent a self-consistent physical scenario are shown in lighter shading. This issue is relevant for the heavier final-state fermions, taus and $b$-quarks, and particularly acute for taus. Finally note that the irregular shape of the contours for the one-step electrons and muons can be traced to the fact the 0-step FSR spectrum is both sharply peaked and has a kinematic edge, leading to a poor fit.

In Fig.~\ref{fig:BestFit} we show the $\Delta \chi^2$ values between the best fit at a given step number $n$ and the best fit overall, for each final state. We show results for both $\epsilon_f=0.3$ and $0.1$ in all cases, and also include $\epsilon_f=0.01$ for electrons. As expected the results do not depend strongly on $\epsilon_f$, especially in the case of taus, which is in accord with the results of Fig.~\ref{fig:0step}. Note that the nominal overall best fit for the taus ($n=4$) falls into the kinematically disallowed (inconsistent) region; $n=4$ cannot be physically accommodated within 3$\sigma$ of its preferred DM mass. For this reason the results for taus and $b$-quarks were rerun allowing only self-consistent scenarios (in the sense of Eq.~\ref{eq:kinematic}); in these cases we obtain the results shown by the blue dotted curves in Fig.~\ref{fig:BestFit}. We summarize the best fit results for $\epsilon_f=0.3$ in Table~\ref{table:results0p3} and the 1$\sigma$ range as determined from Fig.~\ref{fig:BestFit} on these parameters in Table~\ref{table:results1sig}. 

\begin{table}[h]\vspace{0.18in}
\begin{center}
\begin{tabular}{| c || c | c | c | c |}
    \hline
    Final State & $n$-step & $m_\chi$ (GeV)& $\sigma v$ ($\textrm{cm}^3/\textrm{sec}$) & $\chi^2$ \\ \hline \hline
    e & 5 & 67.2 & $2.9 \times 10^{-24}$ & 26.82 \\ \hline
    $\mu$ & 4 & 53.0 & $9.9 \times 10^{-25}$ & 26.94 \\ \hline
    $\tau_{\textrm{unphysical}}$ & 4 & 59.4 & $4.6 \times 10^{-26}$ & 24.13 \\ \hline
    $\tau_\textrm{physical}$ & 2 & 24.1 & $1.4 \times 10^{-26}$ & 25.59 \\ \hline
    $b$ & 2 &  91.2 & $ 3.9 \times 10^{-26}$ & 22.42 \\
    \hline
\end{tabular}
\end{center}
\caption{Best fit to DM annihilations to various final states with $\epsilon_f = 0.3$. For the case of taus we show a best fit point if we include kinematically disallowed masses (unphysical) and also if we restrict ourselves to physical masses as discussed in Sec.~\ref{sec:methods}. Fits were performed over 20 degrees of freedom.}
\label{table:results0p3}
\end{table}

\begin{table}[h]\vspace{0.2in}
\begin{center}
  \begin{tabular}{| c || c | c | c |}
    \hline
    Final State & $n$-step & $m_\chi$ (GeV)& $\sigma v$ ($\textrm{cm}^3/\textrm{sec}$) \\ \hline \hline
    e & 3-6 & 28-107 & $10^{-24.0}$-$10^{-23.3}$ \\ \hline
    $\mu$ &  2-5  & 22-89  & $10^{-24.5}$-$10^{-23.7}$ \\ \hline
    $\tau_{\textrm{unphysical}}$ & 3-5  & 37-94 & $10^{-25.6}$-$10^{-25.1}$ \\ \hline
    $\tau_\textrm{physical}$ & 2 & 24.1 & $10^{-25.8}$ \\ \hline
    $b$ & 0-3 & 40-150  & $10^{-25.8}$-$10^{-25.2} $ \\
    \hline
  \end{tabular}
\end{center}
\caption{Range of parameters within 1$\sigma$ of the best fit step for $\epsilon_f=0.3$ for electrons, muons, taus and $b$-quarks. As in Table~\ref{table:results0p3} we show both physical and unphysical tau results.}
\label{table:results1sig}
\end{table}

\begin{figure}[t!]
\centering
\includegraphics[scale=0.7]{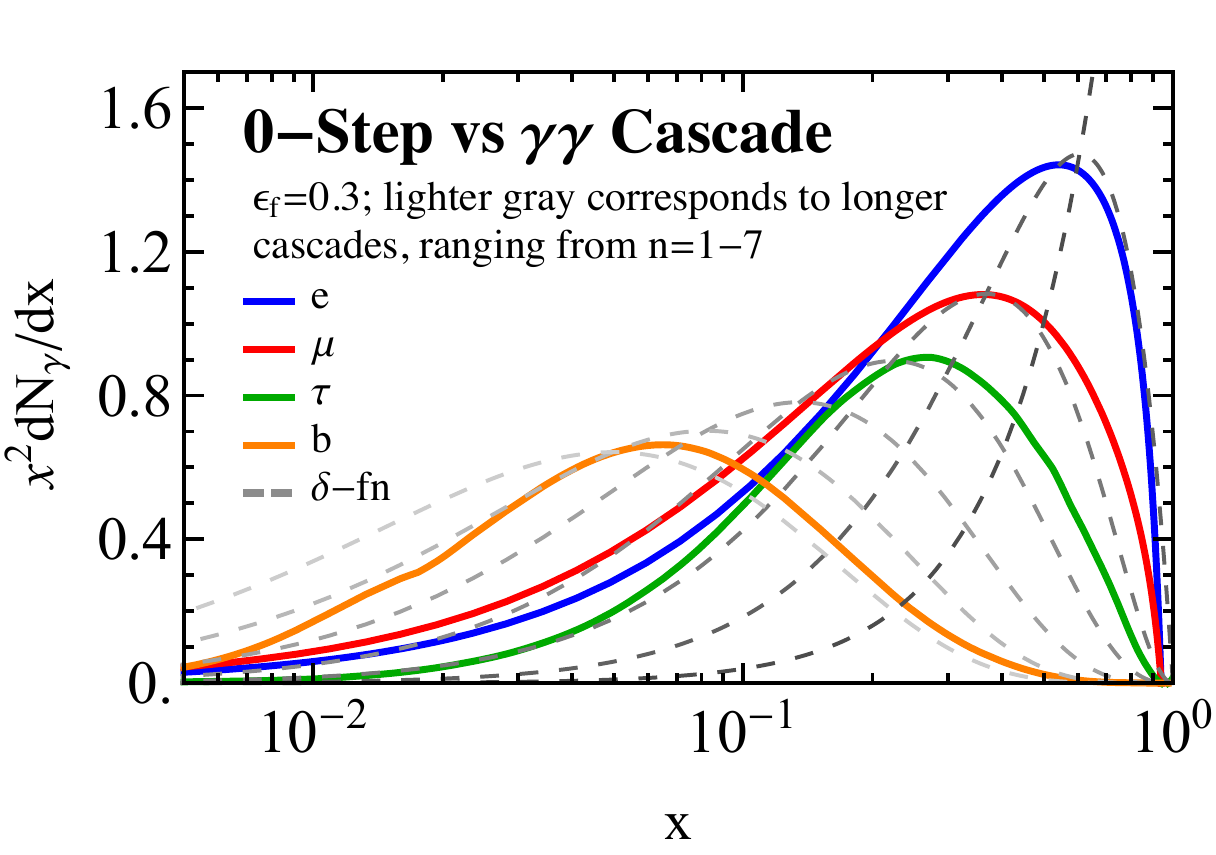}
\caption{\footnotesize{The 0-step spectra for e, $\mu$, $\tau$ and $b$-quarks with $\epsilon_f=0.3$ are shown as the blue, red, green and orange curves. The dashed curves show the spectrum of a hierarchical $n$-step cascade that ends in $\phi_1 \to \gamma \gamma$ (a $\delta$-function in the $\phi_1$ rest frame) for $n=1-7$, with lighter curves corresponding to progressively longer cascades. In order to compare the shape of the spectra we have magnified the 0-step spectra by a factor of $470$, $190$, $6.2$ and $3.1$ for e, $\mu$, $\tau$ and $b$-quarks respectively. We see the electron spectrum is closest to a 2-3 step cascade ending in a $\delta$-function, muons and taus are closest to a 3-4 step cascade, whilst $b$-quarks most resemble 6-7.}}
\label{fig:DeltaCascade}
\end{figure}

In Fig.~\ref{fig:Spectrum0p3Full} we show the overall best fit spectrum for electron, muons, taus, and $b$-quarks with $\epsilon_{f} = 0.3$. Although the spectra for direct annihilation to these final states are quite different, after introducing the freedom to have multi-step cascades, a similar best fit spectrum is picked out in each case. 
To expand on this, we can compare the various 0-step spectra - as displayed in Fig.~\ref{fig:0step} - to the result of a hierarchical $n$-step cascade that ends in $\phi_1 \to \gamma \gamma$. This comparison is shown in Fig.~\ref{fig:DeltaCascade}. The spectrum of photons from this process is just a $\delta$-function in the $\phi_1$ rest frame, and is in a sense the simplest possible photon spectrum. We find that the photon spectrum from direct annihilation to electrons is similar to that obtained by a 2-3 step cascade terminating in $\phi_1 \rightarrow \gamma \gamma$; for muons and taus the closest match is a 3-4 step cascade; and for $b$-quarks 6-7. Of course these correspondences are not exact -- for example, the $b$-quark spectrum is more complex than just applying Eq.~\ref{eq:boosteq} to a $\delta$-function -- but they allow us to regard these 0-step spectra as arising approximately from a common ($\delta$-function) spectrum convolved with differing numbers of cascade steps. We can then intuit how many additional steps are required in each case, to bring the spectra to a similar shape. Combining these numbers with the preferred number of steps seen in Table~\ref{table:results0p3}, we find the GCE prefers a spectrum that can be roughly modeled as a $\delta$-function occurring at the endpoint of 7-9 cascade decays. In this sense it seems fits to the GCE prefer a cascade with a large number of steps, and that these can occur in the SM or dark sector.

Likewise, this general picture can approximately describe showers in the dark sector \cite{Freytsis:2014sua}. Such showers will effectively contain decay cascades of different lengths, but we find that the spectrum of \cite{Freytsis:2014sua} can be well described by a $\delta$-function $\phi_1 \rightarrow \gamma \gamma$ broadened by $\sim 3$ decay steps. The best-fit scenario found in that paper corresponds to a DM mass of $\sim 10$ GeV; this is consistent with the preferred mass for our 1-step electron case, which also corresponds to a $\delta$-function at the endpoint of a $\sim 3$-step cascade. A better fit to the data might therefore be obtained by combining such dark showering with a short dark-sector cascade. In Sec.~\ref{sec:generalcascade} we will return to this point, and discuss the sense in
which our results may be used to estimate the parameter space for dark shower models.

\subsection{Different Final States}
A few comments about the various final states are in order.

\textit{Electrons:} The photon spectrum from direct annihilations $\chi \chi \rightarrow e^+ e^-$ is sharply peaked. This tends to produce a worse fit to the GCE. As such we need several steps in the cascade in order to broaden the spectrum sufficiently to allow for a parameter space where a significantly improved fit is possible, and this is shown by the substantial decrease in the quality of fit at low $n$ in Fig.~\ref{fig:BestFit}. It should be noted that any model for the GCE with direct annihilation into electrons will likely be in severe tension with the data from AMS \cite{Bergstrom:2013jra}. This tension is likely to persist for at least the $n=1$ cascade, and possibly higher steps as well \cite{Cline:2015qha}. As we go to higher-step cascades the spectrum broadens and the AMS bounds are expected to weaken, but the exact bounds should be worked out for any cascade scenario with a branching fraction to electrons. For the purposes of this work, we use the electron case as an example of a sharply peaked photon spectrum to demonstrate the impact of the spectral broadening, not necessarily as a realistic explanation for the excess. Similarly, constraints on DM annihilation from the cosmic microwave background (CMB) \cite{Planck:2015xua} are likely to rule out both the electron and muon favored regions shown in Fig.~\ref{fig:AllStatesLogChiPlot}, while leaving the $b$ and tau regions largely unconstrained. The figure of merit for CMB constraints is $\langle \sigma v \rangle/m_\chi$ \cite{Chen:2003gz, Padmanabhan:2005es}, up to an $\mathcal{O}(1)$ factor which is channel- and spectrum-dependent \cite{Slatyer:2009yq, Madhavacheril:2013cna}. As discussed above, for the best-fit regions (for hierarchical decays), this quantity scales as $\sim m_\chi^{0.3}$ as the number of steps increases; thus, we expect the constraint to become slightly stronger for longer cascades.

\textit{Muons:} In Fig.~\ref{fig:BestFit} we see that the muon final state spectrum has the same qualitative behavior as the electrons, and will be subject to similar constraints. This is unsurprising as the muon spectrum is quite similar to that from electrons, albeit with a less pronounced peak (see Fig.~\ref{fig:0step}).

\textit{Taus:} As with other leptonic final states, taus also prefer multi-step cascades for the best fit. Note that the best fit point at 4 steps is in fact kinematically disallowed (inconsistent) as can be seen in Fig.~\ref{fig:AllStatesLogChiPlot} and as discussed in Sec.~\ref{sec:methods}. However, the best fit point after imposing the consistency condition, at 2 steps, is still a better fit than the high-step cases with electron and muon final states.

\textit{$b$-quarks:} DM annihilation to $b$-quarks is the preferred channel for direct annihilation identified in \cite{2014arXiv1402.6703D,Calore:2014xka}, where it already provides a good fit. Accordingly there is no need to broaden the spectrum with a large number of cascades -- however, as we will discuss in Sec.~\ref{sec:signalsconstraints}, even a short cascade can greatly alleviate constraints from colliders and direct searches (see also \cite{Martin:2014sxa,Abdullah:2014lla} and references therein). A cascade with several steps can still give an equally good or slightly better fit, and of course accommodates higher masses than for the case of direct annihilation. However, since the spectrum is already fairly broad, adding too many additional steps makes the fit worse, as shown in Fig.~\ref{fig:BestFit}. Accordingly, the DM mass cannot be pushed far above 100 GeV without significantly worsening the fit, at least in the context of hierarchical cascades.

\subsection{Sensitivity to Systematics and Energy Cuts}

\begin{figure}[t!]
\centering
\includegraphics[scale=0.65]{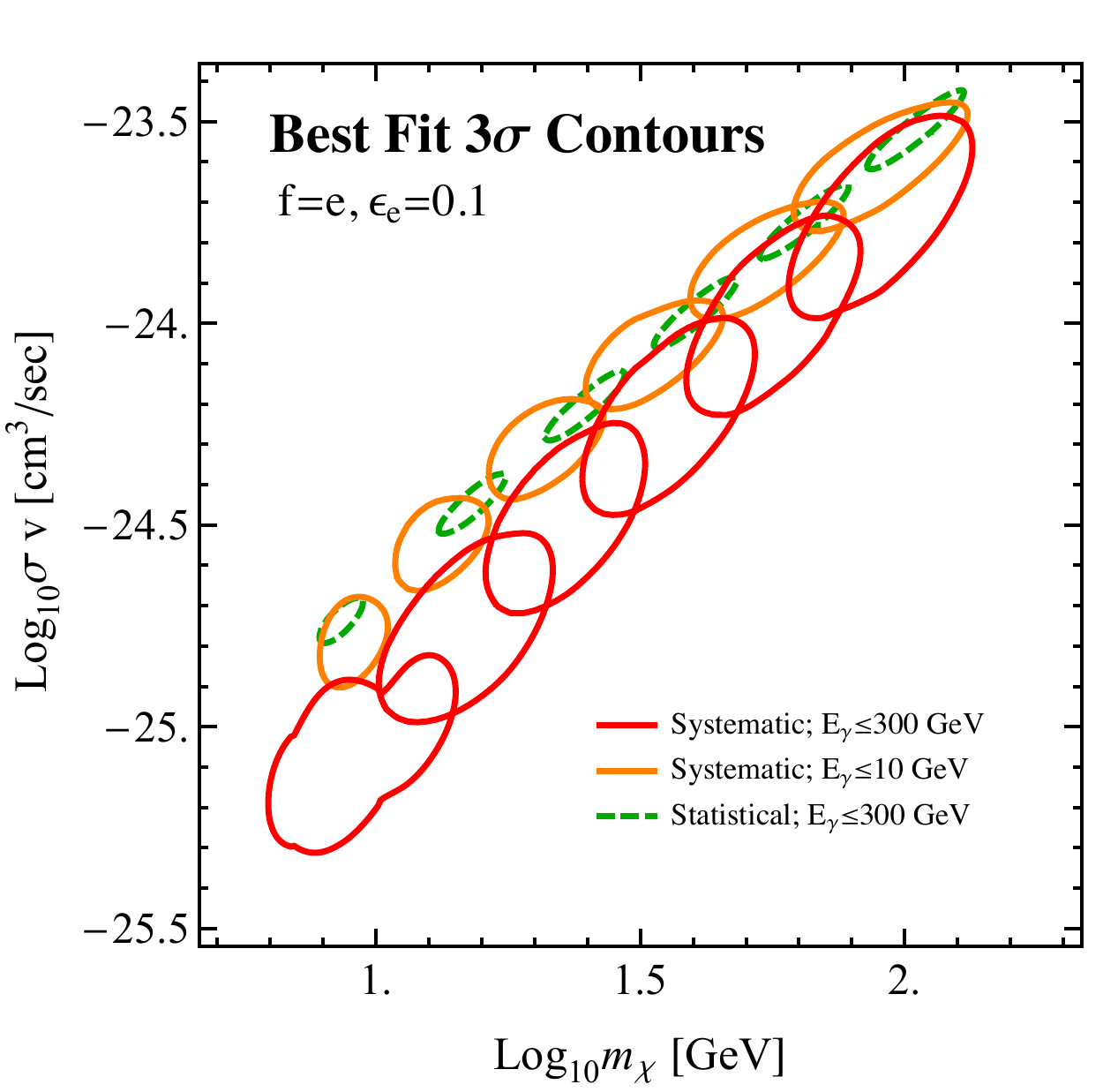}
\caption{\footnotesize{The 3$\sigma$ contours for 1-6 step cascade annihilations to final state electrons with $\epsilon_e = 0.1$. Red contours correspond to fitting over the entire energy range $0.5~\textrm{GeV} \leq E_\gamma \leq 300~\textrm{GeV}$ with the full covariance matrix of \cite{Calore:2014xka}. Orange contours correspond to fitting with a cut on high energies $E_\gamma \leq 10~\textrm{GeV}$. Green contours correspond to a fit over the full energy range but with only the statistical errors of \cite{Calore:2014xka}.}}
\label{fig:Electron0p1StatsPlot}
\end{figure}

\begin{figure*}[t!]
\centering
\begin{tabular}{c}
\hspace{0.04in}\includegraphics[scale=0.6]{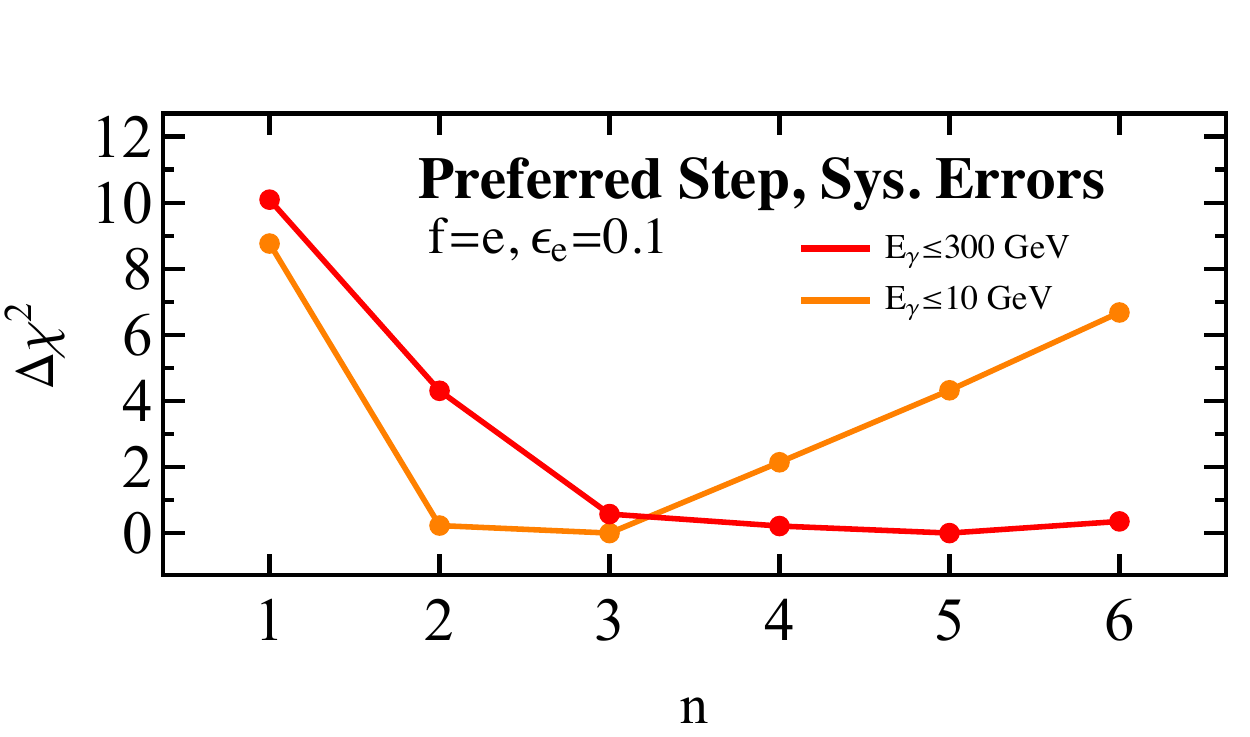} \hspace{0.16in}
\includegraphics[scale=0.61]{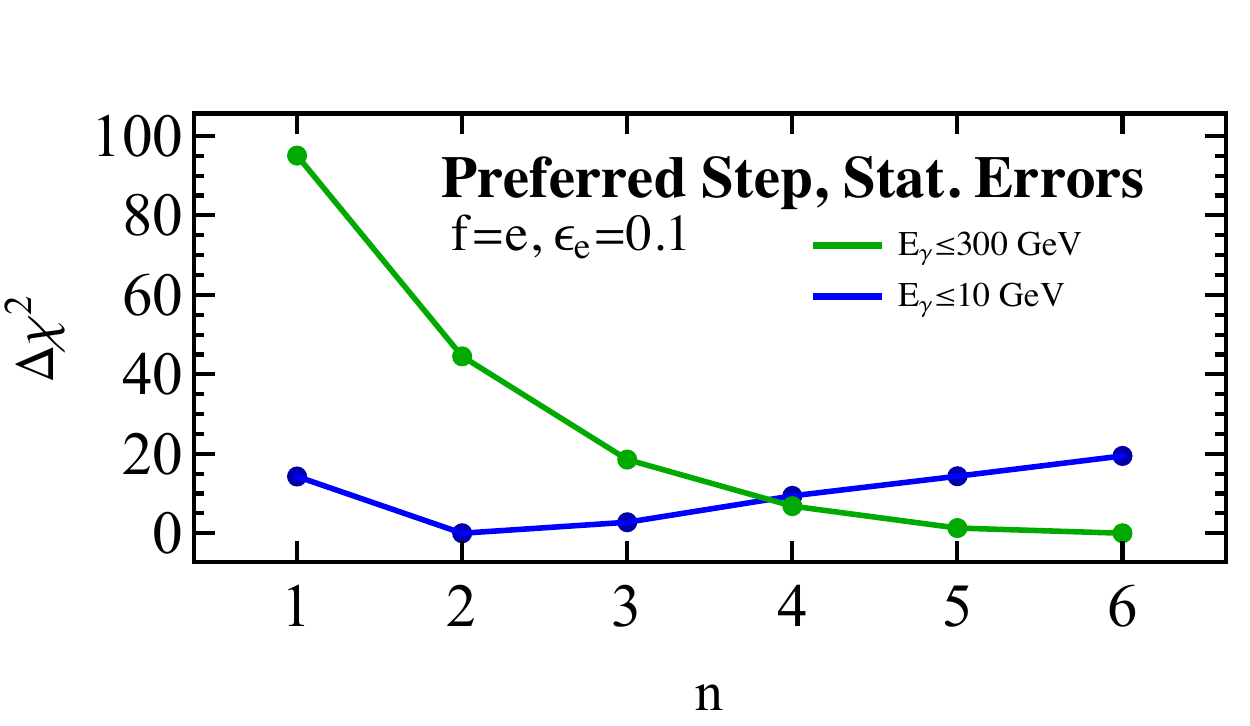} \\
\includegraphics[scale=0.635]{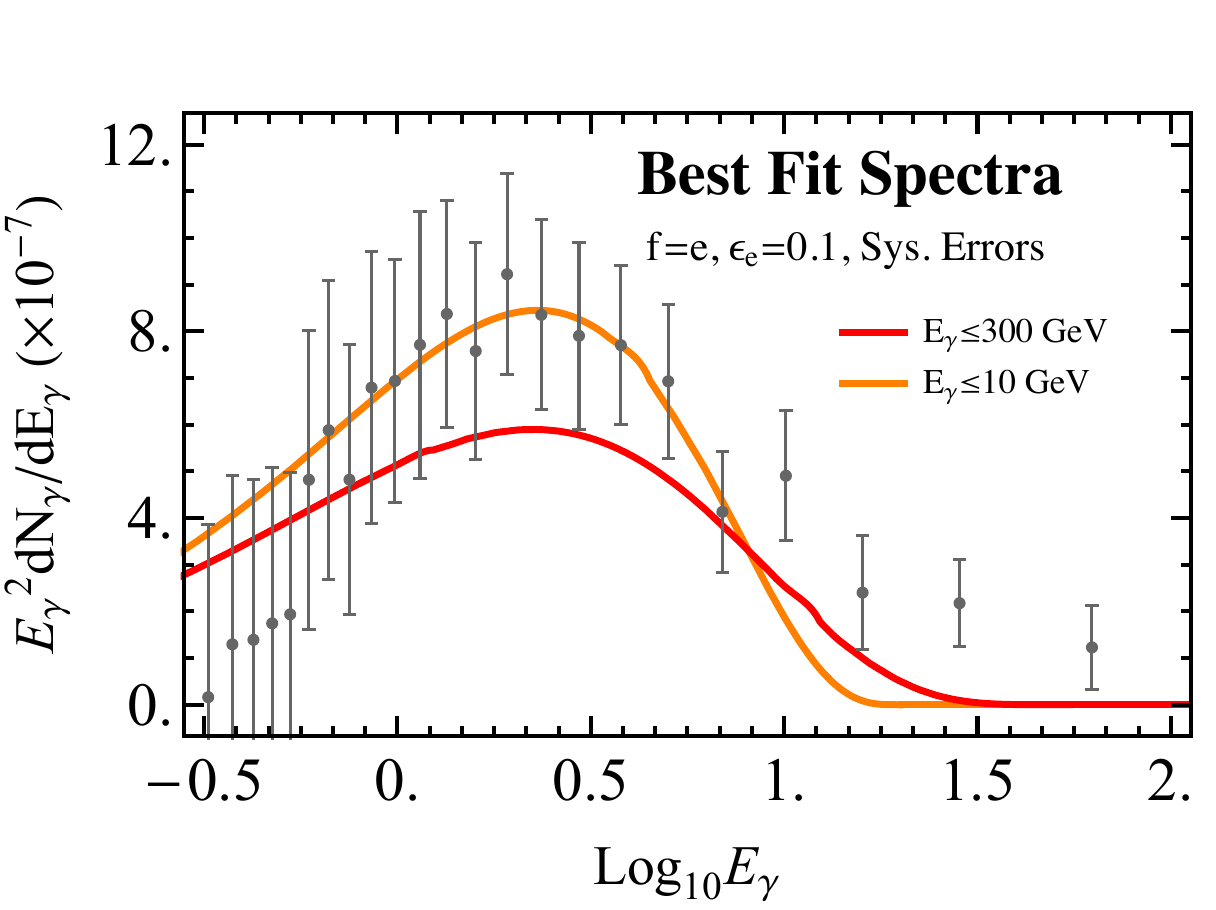} \hspace{0.15in}
\includegraphics[scale=0.635]{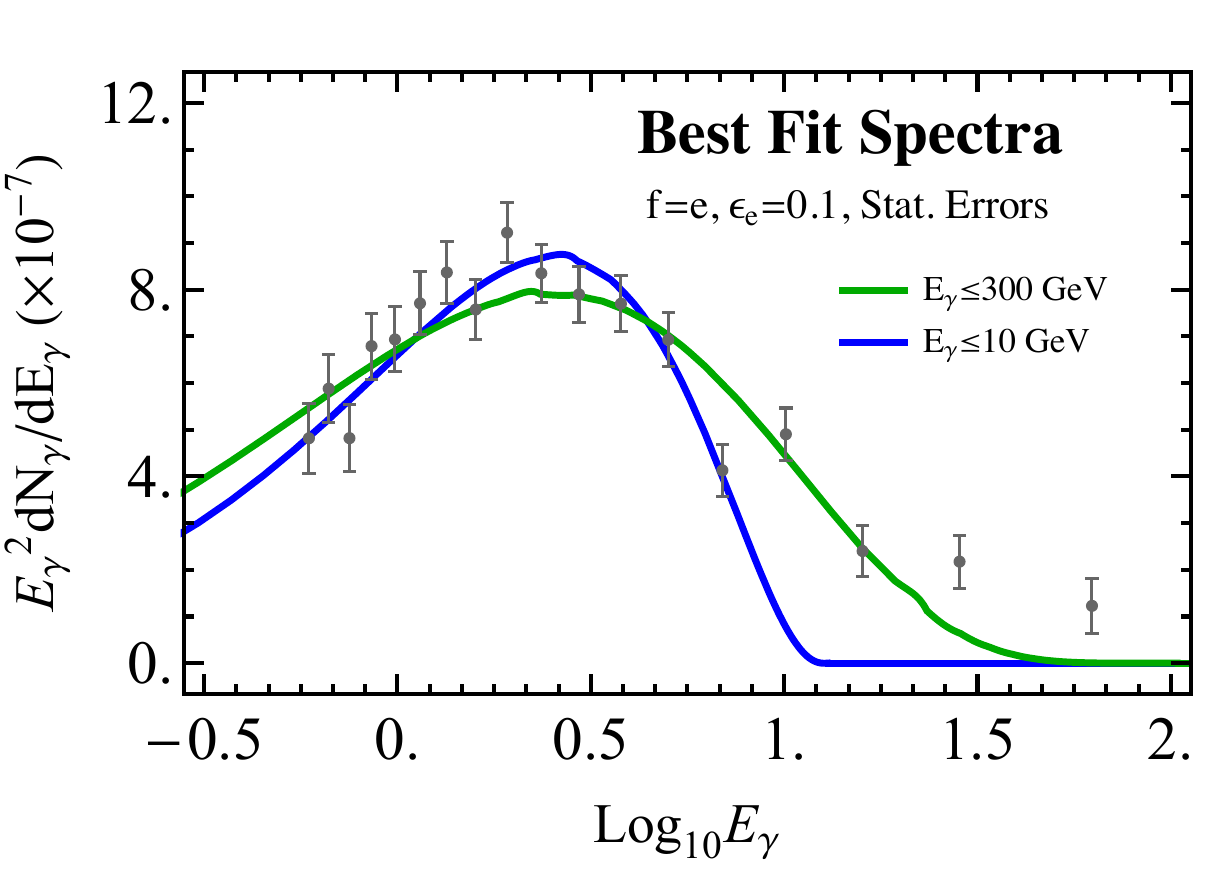}\\
\end{tabular}
\caption{\footnotesize{Top Panels: We show the impact on the preferred number of steps when changing the energy range and error types considered. Each curve is for final state electrons with $\epsilon_e=0.1$. The left figure shows the use of systematic errors over the full and a restricted energy range ($E_\gamma \leq 10$ GeV) in red and orange respectively. The right figure is the equivalent for statistical errors, with the full energy range shown in green and the restricted in blue. Bottom panels: Here the best fit curves as determined from the top panels are shown with the appropriate data and errors from \cite{Calore:2014xka} overlaid, for the example case of the electron final state. The left panel shows the results for systematic errors, where the best fit point was $n=5$ for the full range (red curve) and $n=3$ for the restricted range (orange curve). The right panel shows the equivalent for statistical errors, where for the full range the $n=6$ curve is shown in green and for the restricted range the $n=2$ curve is in blue.}}
\label{fig:StatsDeltaChiandSpecPlot}
\end{figure*}

In the results presented above we have fit the data of \cite{Calore:2014xka} to the photon spectrum from DM annihilations through multi-step cascades to various final states. The fit was performed over the energy range $0.5~\textrm{GeV} \leq E_{\gamma} \leq 300~\textrm{GeV}$. There is some evidence that the emission detected above 10 GeV may not share the same spatial profile as the main excess, suggesting a possible independent origin (for example, these high-energy data appear to prefer a morphology centered at negative $\ell$ and with a shallow spatial slope \cite{Calore:2014xka}), so we also test the impact of omitting the data above 10 GeV. Finally, we explore the impact of including only the statistical uncertainties of \cite{Calore:2014xka}, omitting systematic errors, to test the degree to which the constraints could improve with reduction in the systematic uncertainties.

We display the results of this study in Fig.~\ref{fig:Electron0p1StatsPlot}-\ref{fig:StatsDeltaChiandSpecPlot}, for the case of $n$-step cascade annihilations to final state electrons with $\epsilon_e = 0.1$. Annihilations to other final states generically display the same behavior as the energy range and error estimates are varied. Cutting out the high energy data points generically shifts the fit to prefer lower masses and narrower spectra, and therefore corresponds to cascades with fewer steps -- resembling a $\delta$-function at the endpoint of a 5-7 step cascade, rather than a 7-9 step cascade. At a fixed number of steps, the main impact of omitting the high-energy data points is to raise the preferred cross-section and shrink the contours. Understanding the high-energy data will thus be important in distinguishing quantitative models for the GeV excess. 

Fitting over statistical errors increases the actual $\chi^2$ values, and the rate at which $\chi^2$ increases away from its minimum (as expected), as demonstrated by the shrinking green contours of Fig.~\ref{fig:Electron0p1StatsPlot}. The overall preferred step in the cascade however is not dramatically affected, only changing by 0-1 steps, as shown in the top panels of Fig.~\ref{fig:StatsDeltaChiandSpecPlot} - we display the corresponding best fit spectra in the bottom panels. At a fixed number of steps, the preferred cross-section increases, becoming more similar to what we find when omitting the high energy points.

\section{Interpretation for General Cascades}
\label{sec:generalcascade}

\subsection{Relaxing the Assumption of Large Hierarchies}

The results displayed in the previous section were obtained assuming large mass hierarchies between each cascade step.  It is possible to recast these results to gain insight into the case of general $\epsilon_i$ values. To see this, consider the decay $\phi_{i+1} \rightarrow \phi_i \phi_i$. As previously discussed, in the limit when two mass scales become degenerate ($\epsilon_i \to 1$), an $n$-step cascade effectively reduces to an $(n-1)$-step cascade, except for the additional final state multiplicity.  Thus adding a degenerate step to a cascade is much simpler than adding one with a large hierarchy: we need only multiply the spectrum by two to account for the increased multiplicity, and halve the photon energy scale to account for the initial energy being spread between twice as many particles. (For completeness, we check analytically that the limit of $\epsilon_i\rightarrow 1$ has this behavior in Appendix~\ref{app:boost}.)

\begin{figure}[t!]
\centering
\includegraphics[scale=0.65]{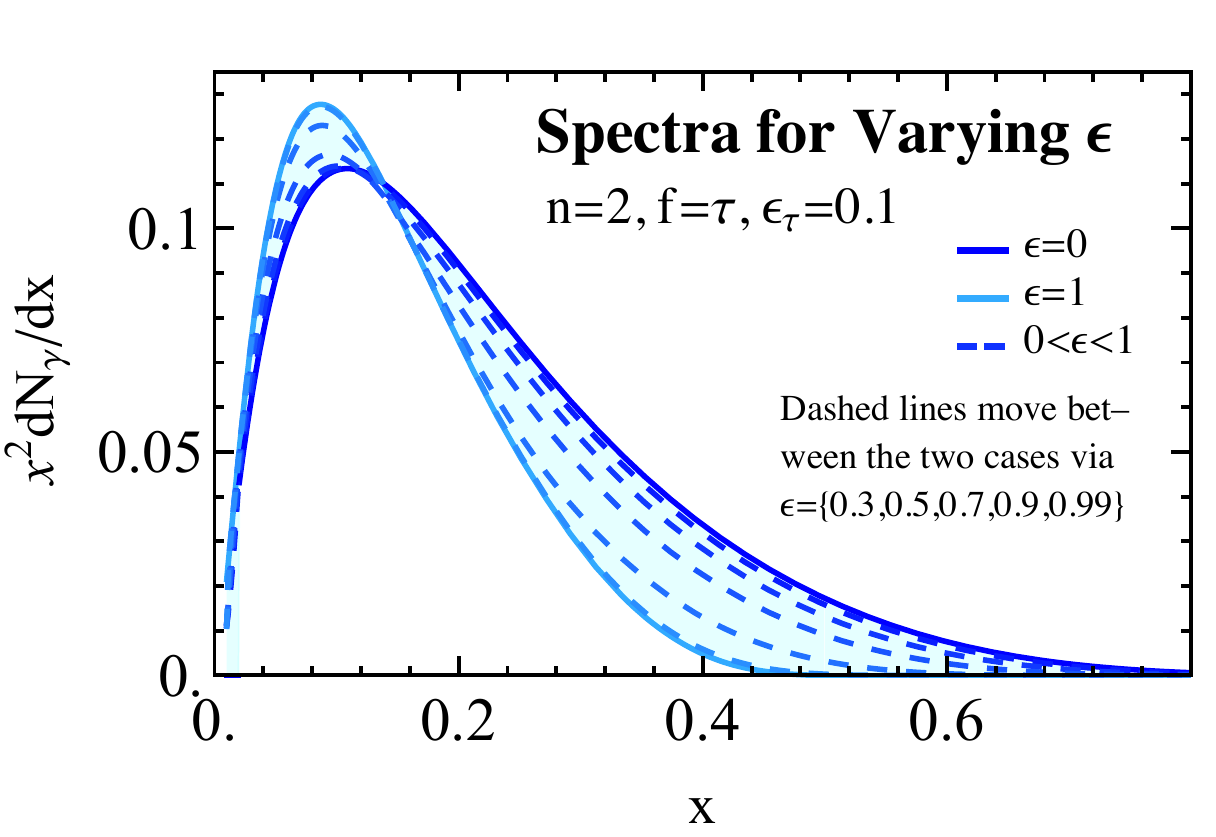}
\caption{\footnotesize{The transition of the spectra between $\epsilon_2=0$ and $\epsilon_2=1$, calculated using Eq.~\ref{eq:fullepsilon}. The example case is a 2-step cascade with final state taus and $\epsilon_\tau = 0.1$. The dark blue is for $\epsilon=0$ and is what would result from the large hierarchies approximation. The $\epsilon=1$ case shown in light blue corresponds to a completely degenerate spectrum, and as such is equivalent to a shifted 1-step curve. In between these two, we show intermediate $\epsilon$ values as the dashed curves, specifically $\epsilon=\{0.3,0.5,0.7,0.9,0.99\}$. Note the rate of transition between the two cases is in keeping with the error in the large hierarchies case being of order $\mathcal{O}(\epsilon_i^2)$.}}
\label{fig:fullEpsilon0p1}
\end{figure}

In light of this, an $n$-step cascade with one degenerate step and an $(n-1)$-step hierarchical cascade must provide equally good fits to the GCE, with the former preferring twice the annihilation cross-section and DM mass relative to the latter. The increased DM mass results from the halving of the energy scale, whilst to understand the cross-section we look back to Eq.~\ref{eq:xsecnorm}: adding the degenerate step doubles the photon multiplicity, which halves $\eta$ to compensate, but the doubling of the DM mass means overall the cross-section is increased by a factor of two. As such the results in Fig.~\ref{fig:AllStatesLogChiPlot} can be readily extended for additional degenerate steps. For each additional degenerate step on top of an initial hierarchical cascade (the degenerate step may occur anywhere in the cascade), the shape of the $\chi^2$ contours remains the same, but shifted upward by a factor of two in mass and cross-section. With a sufficiently large number of degenerate decays, the DM mass required to fit the GCE could be made arbitrarily high, although this would seem to require considerable fine-tuning. (A natural scenario in which one degenerate step arises due to a symmetry is discussed in \cite{Fan:2012gr}.)

\begin{figure}[t!]
\centering
\includegraphics[scale=0.6]{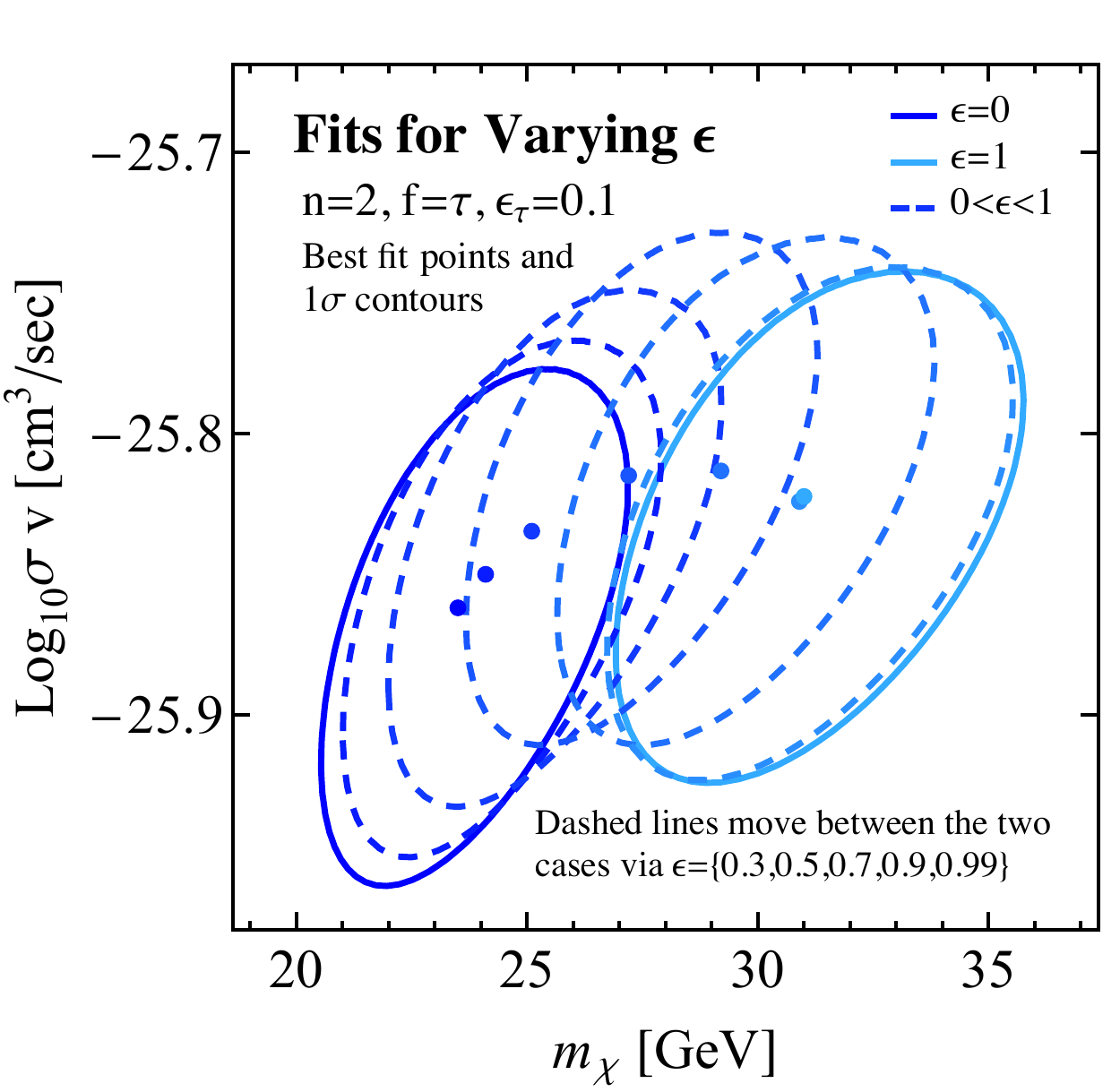}
\caption{\footnotesize{The transition of the best fit point and 1$\sigma$ contours between $\epsilon_2=0$ and $\epsilon_2=1$, calculated using Eq.~\ref{eq:fullepsilon}. The example case is a 2-step cascade with final state taus and $\epsilon_\tau = 0.1$. The transition is between the $\epsilon=0$ case in dark blue and $\epsilon=1$ in light blue. The dashed curves map out the transition with intermediate values, specifically $\epsilon=\{0.3,0.5,0.7,0.9,0.99\}$.}}
\label{fig:fullEpsilonFits}
\end{figure}

Cascades with general values of $\epsilon_i$ in turn interpolate between the two simpler cases already considered, with small and large $\epsilon_i$. We give the general convolution formula in Appendix~\ref{app:boost}, and an example of how spectra evolve as a single $\epsilon_i$ shifts from 0 to 1 is shown in Fig.~\ref{fig:fullEpsilon0p1}. This interpolation provides an alternate interpretation for Fig.~\ref{fig:BestFit}: the $n$ on the $x$-axis of these plots can be thought of as representing the number of steps with a large hierarchy, rather than the total number of steps. If one of these steps becomes degenerate (while holding the total number of steps fixed), as previously discussed, we will move from $n$ to $n-1$ steps in terms of the spectral shape and hence quality of fit. Intermediate $\epsilon_i$ values will interpolate smoothly between these two cases. Thus for any arbitrary collection of hierarchical and degenerate steps, the quality of the fit and the location of the best-fit region in $m_\chi-\langle\sigma v\rangle$ parameter space can already be estimated from Figs.~\ref{fig:AllStatesLogChiPlot}-\ref{fig:BestFit}. A concrete example of the transition in preferred DM mass and cross-section is shown in Fig.~\ref{fig:fullEpsilonFits}, which corresponds to the variation of the spectrum shown in Fig.~\ref{fig:fullEpsilon0p1}. The curve plotted out by the best fit point for intermediate values of $\epsilon$ is not a straight line between the two extreme values, but does not deviate far from this. Similar behavior was seen for other final states and choice of degenerate step.

At a fixed DM mass, the perturbation to the $\epsilon_i=0$ photon spectrum evolves roughly as $\epsilon_i^2$ as $\epsilon_i$ varies from 0 to 1 (as discussed in Appendix~\ref{app:boost}); this behavior is shown in Fig.~\ref{fig:fullEpsilon0p1}, where the $\epsilon_2=0.3$ spectrum is almost indistinguishable from the $\epsilon_2=0$ spectrum, and $\epsilon_2=0.5$, $\epsilon_2=0.7$ and $\epsilon_2=0.9$ give spectra intermediate between the $\epsilon_2=0$ and $\epsilon_2=1$ cases. The perturbation to the best-fit $\chi^2$ will tend to increase even \emph{more} slowly than $\epsilon_i^2$, in the case where $\epsilon_i=0$ is a better fit than $\epsilon_i=1$, since the DM mass and cross-section can float to absorb changes in the spectrum and reduce the increase in $\chi^2$. In all examples tested the best-fit $\chi^2$ remains essentially unchanged from the $\epsilon_i=0$ case out to $\epsilon_i=0.7$.

\begin{figure}[t!]
\centering
\includegraphics[scale=0.65]{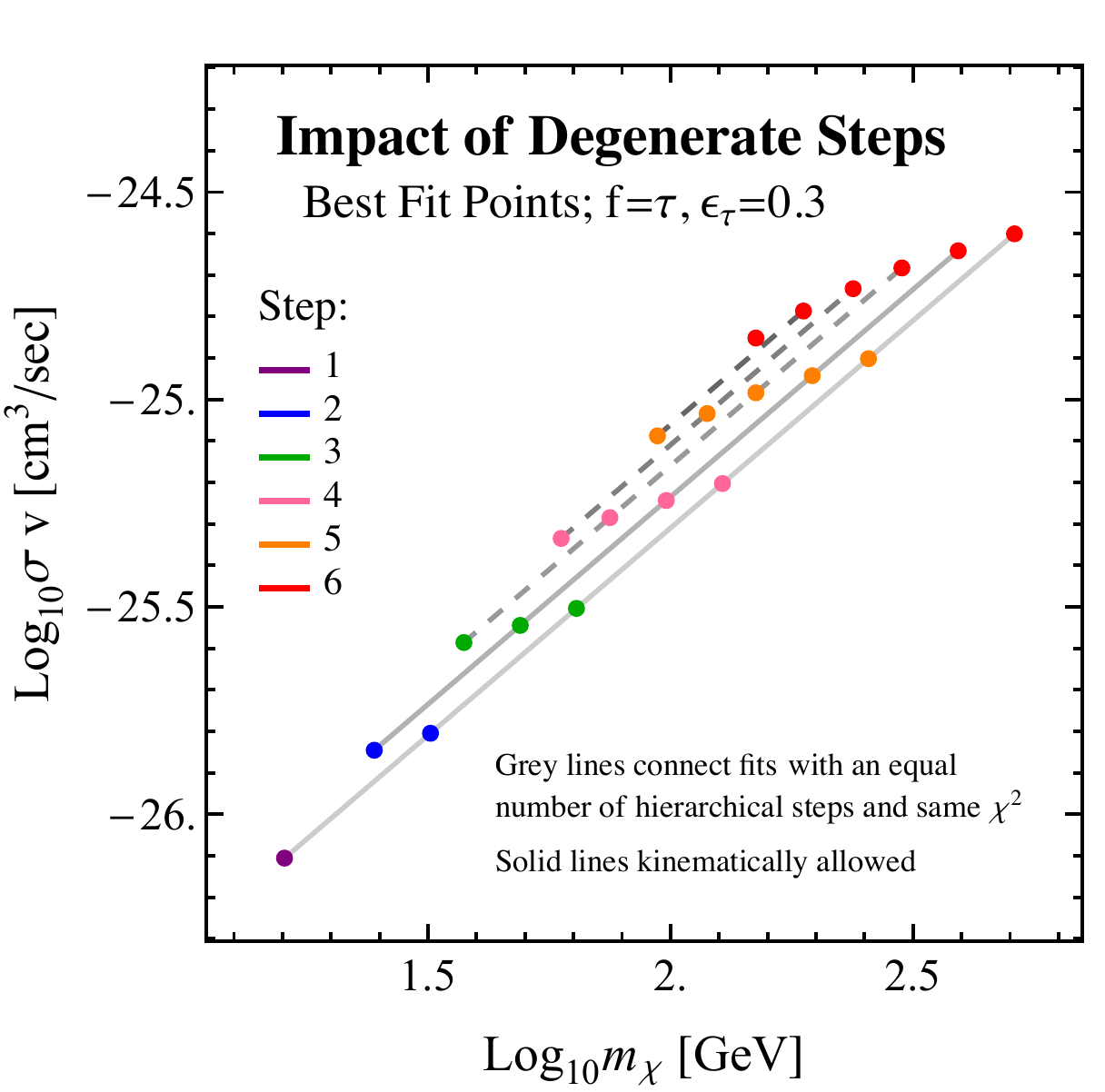}
\caption{\footnotesize{The purple, blue, green, pink, orange and red points correspond to the best fit $(m_{\chi}, \sigma v)$ point for a total number of cascade steps (degenerate + hierarchical) $n$ = 1, 2, 3, 4, 5, 6 respectively; for annihilations to final state taus with $\epsilon_\tau = 0.3$. Points living on the same line have the same number of hierarchical steps and therefore result in equally good fits to the data. Points of the same color, but with progressively greater values of $(m_\chi, \sigma v)$, correspond to successively replacing hierarchical steps with degenerate steps, holding the number of total steps fixed. For the above case of taus only the one and two step hierarchical cascades are kinematically allowed as indicated in Fig.~\ref{fig:AllStatesLogChiPlot} (note that the kinematic constraint applies to lines as a whole, not individual points; see text), thus only points living on the solid lines are allowed as these lines correspond to cascades with one and two hierarchical steps respectively.}}
\label{fig:DegFit}
\end{figure}

In general a cascade with $n$ total steps, $n_d$ of which are degenerate ($n_d$ values of $\epsilon_i \rightarrow 1$) will have the same spectrum as a cascade with $(n-n_d)$ hierarchical steps with a factor of $2^{n_d}$ enhancement in mass and cross-section. This is illustrated in Fig.~\ref{fig:DegFit} for the case of decays to final state $\tau$'s with 1-6 total cascade steps. Relaxing the assumption of large hierarchies therefore results in a preferred triangular slice of parameter space, bounded by curves with $\langle \sigma v \rangle \propto m_\chi$ and $\langle \sigma v \rangle \propto m_\chi^{1.3}$. We can now understand the results of  Fig.~\ref{fig:BestFit}  as mapping out the variation in $\chi^2$ when moving between \emph{classes} of scenarios, each defined by a fixed number of hierarchical steps but containing scenarios with varying numbers of degenerate steps (each of these classes is represented by a line in Fig.~\ref{fig:DegFit}). Note also that the kinematic constraint Eq.~\ref{eq:mDM} acts on classes rather than individual scenarios (since adding a degenerate step doubles the DM mass but increases the number of steps by 1, strengthening the constraint on DM mass by a factor of 2); if one scenario is disallowed the entire class is disallowed.

\begin{figure*}[t!]
\centering
\begin{tabular}{c}
\includegraphics[scale=0.60]{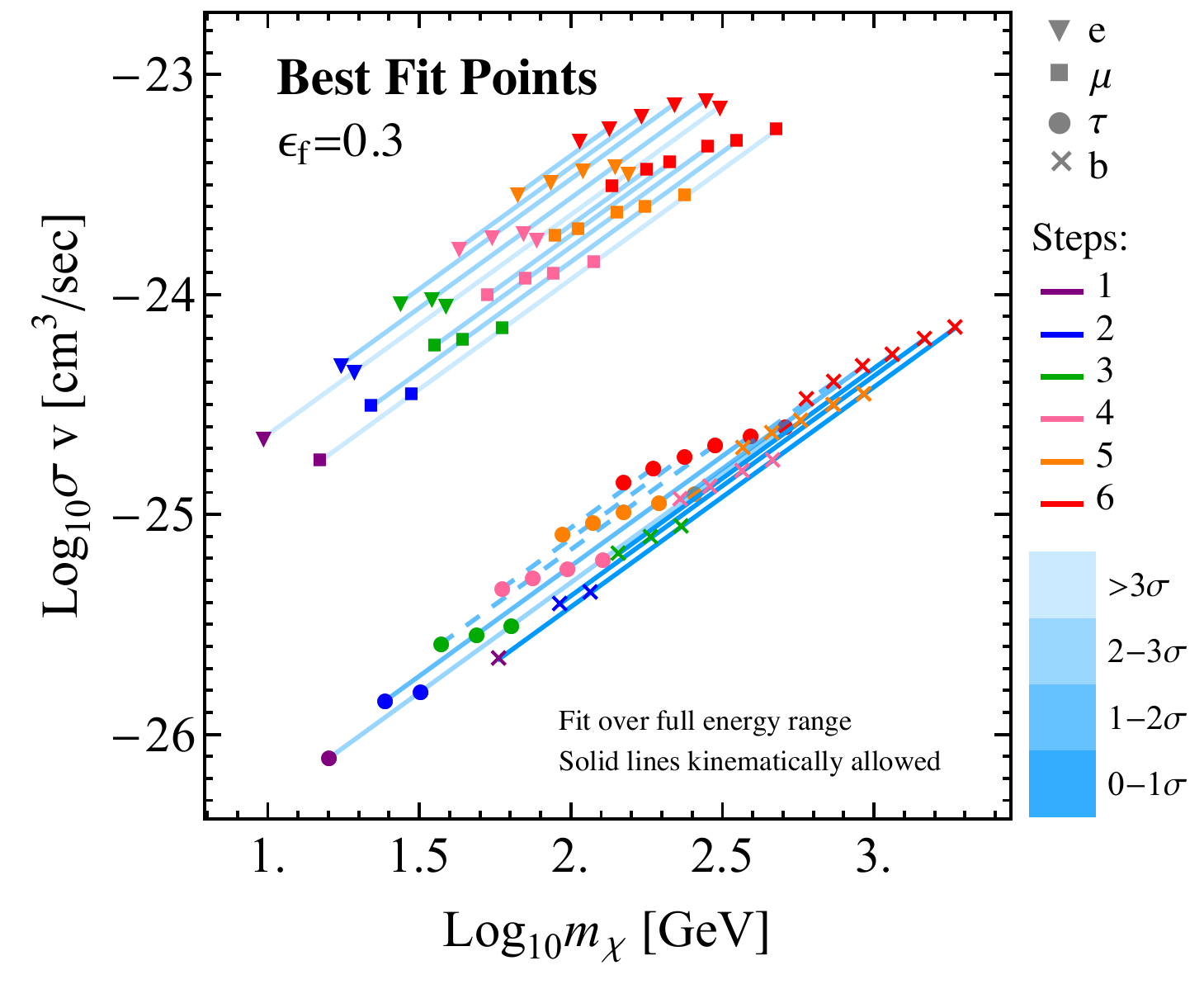}
\includegraphics[scale=0.60]{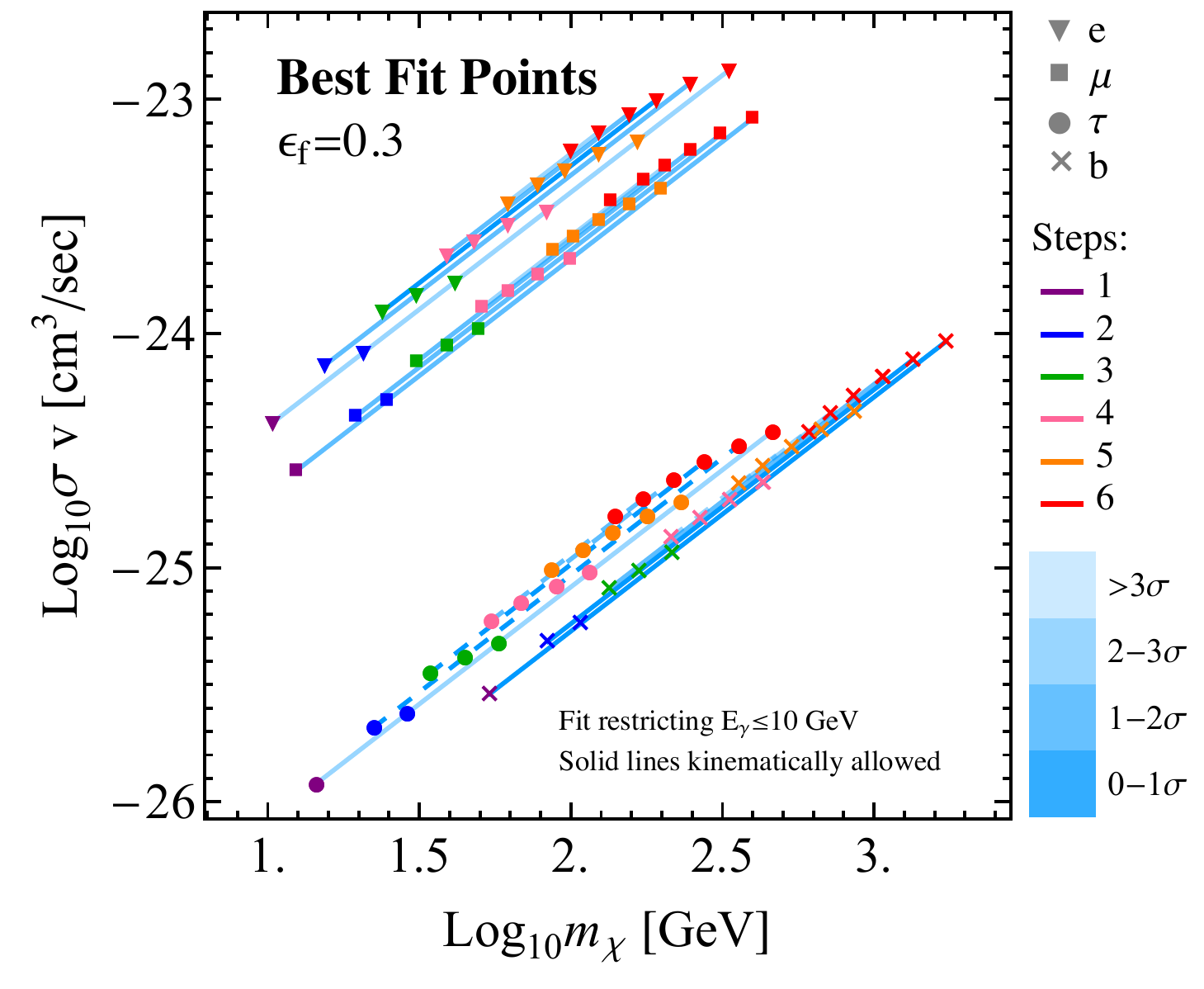}\\
\hspace{0.07in}\includegraphics[scale=0.55]{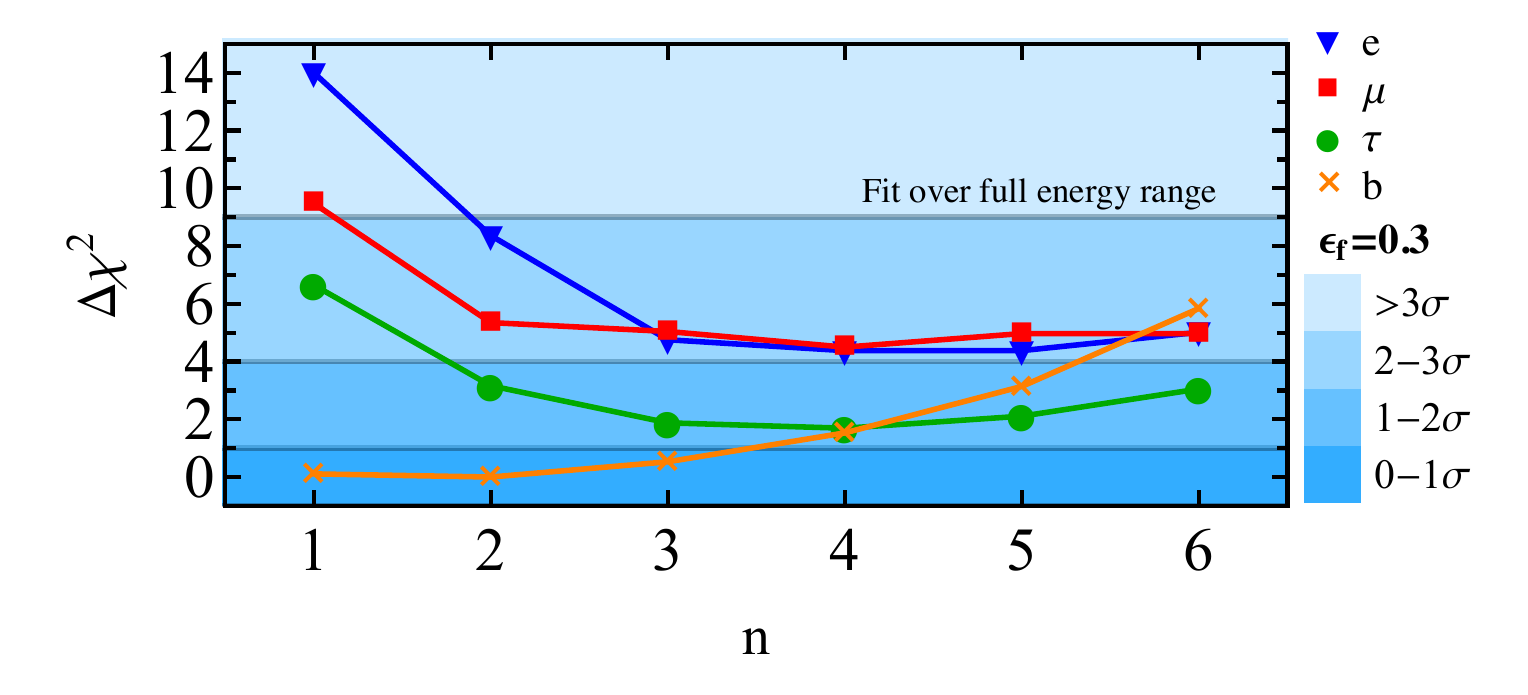}\hspace{0.13in}
\includegraphics[scale=0.55]{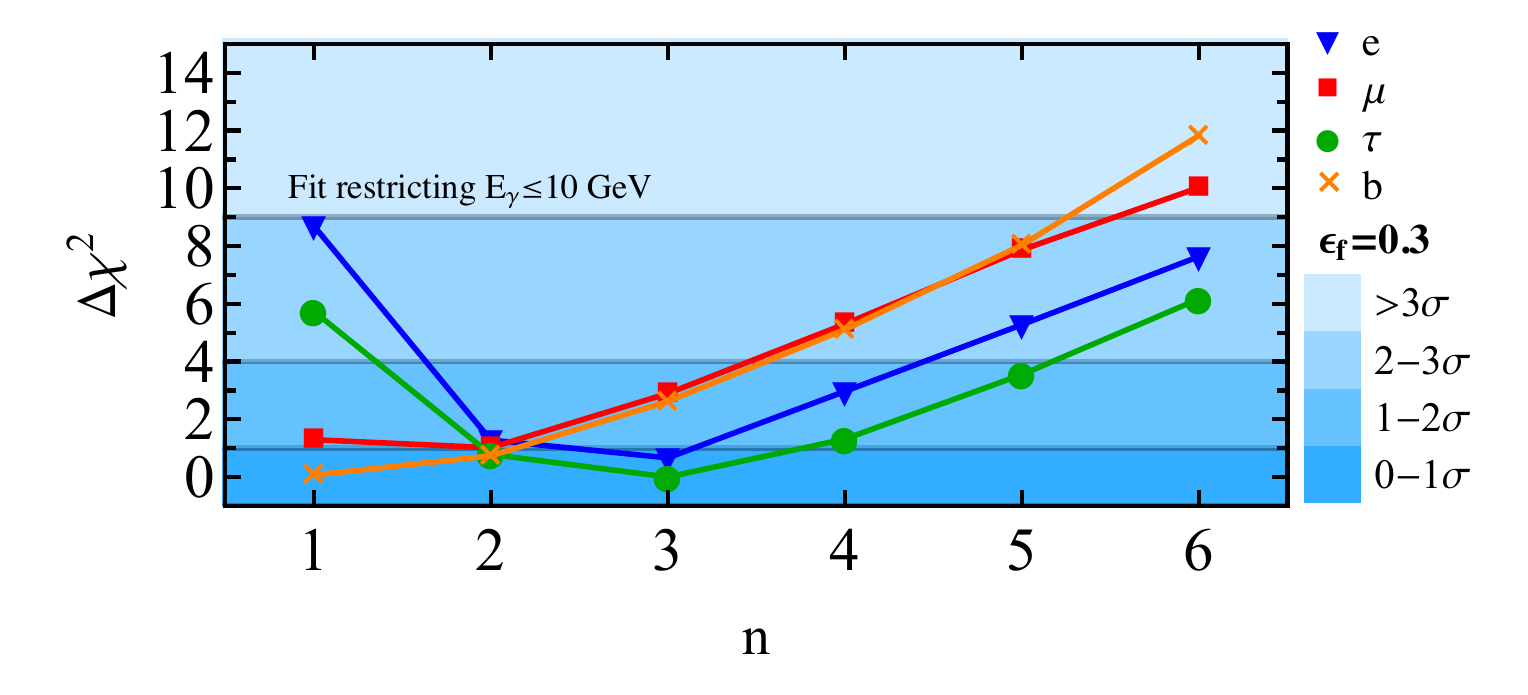}\\
\end{tabular}
\caption{\footnotesize{Combined results of fits with $\epsilon_f = 0.3$ over the full energy range (left) or with a restriction $E_{\gamma} \leq 10$ GeV (right). Top panels: Best fit $(m_{\chi},\sigma v)$ for a cascade with 1-6 total (degenerate + hierarchical) steps ending in electrons, muons, taus of $b$-quarks. Points on the same line have the same number of hierarchical steps and therefore result in equally good fits to the data, following the discussion in Sec.~\ref{sec:generalcascade}. Points of the same color, but with sequentially greater values of $(m_{\chi},\sigma v)$, correspond to progressively replacing hierarchical steps with degenerate steps, holding the total number of steps fixed. The color of the lines indicate goodness of fit and only solid lines are kinematically allowed (as explained in see Sec.~\ref{sec:methods}). Bottom panels: Show the overall best fit for DM annihilation through an $n$-step hierarchical cascade to electron, muon, tau and $b$-quark final states. The curves show the $\Delta \chi^2$ of the best fit at that step and final state, as compared with best fit over all steps and final states. No restriction to physical kinematics is imposed, but where restrictions would apply can be inferred from the top panels. The shaded bands correspond to the quality of fit. For fits over the full energy range a fairly short cascade terminating in decay to $b$-quarks gives the preferred spectrum, whilst over the restricted energy range each final state can potentially provide approximately equally good fits.}}
\label{fig:CombinedResults}
\end{figure*}

Fig.~\ref{fig:CombinedResults} summarizes our combined results. There, the top panels display the regions mapped out in the $\langle \sigma v \rangle-m_\chi$ plane by the best fit points involving 1-6 steps (either hierarchical or degenerate)  cascades to final state electrons, muons, taus and $b$-quarks. In the bottom panels, we indicate which hierarchical step and final state yield the best fit, and the comparative quality of fit for other combinations. We show all these results for fits over the full (left panels) and restricted (right panels) energy ranges. Additionally as shown in the top panels, electrons (taus) and muons ($b$-quarks) have some degree of overlap, especially once degenerate steps are included. The overlap of these regions is reduced when the high energy data points are excluded, as is clear by comparing the right and left panels.

\begin{figure}[t!]
\centering
\vspace{0.15in}
\begin{tabular}{c}
\includegraphics[scale=0.60]{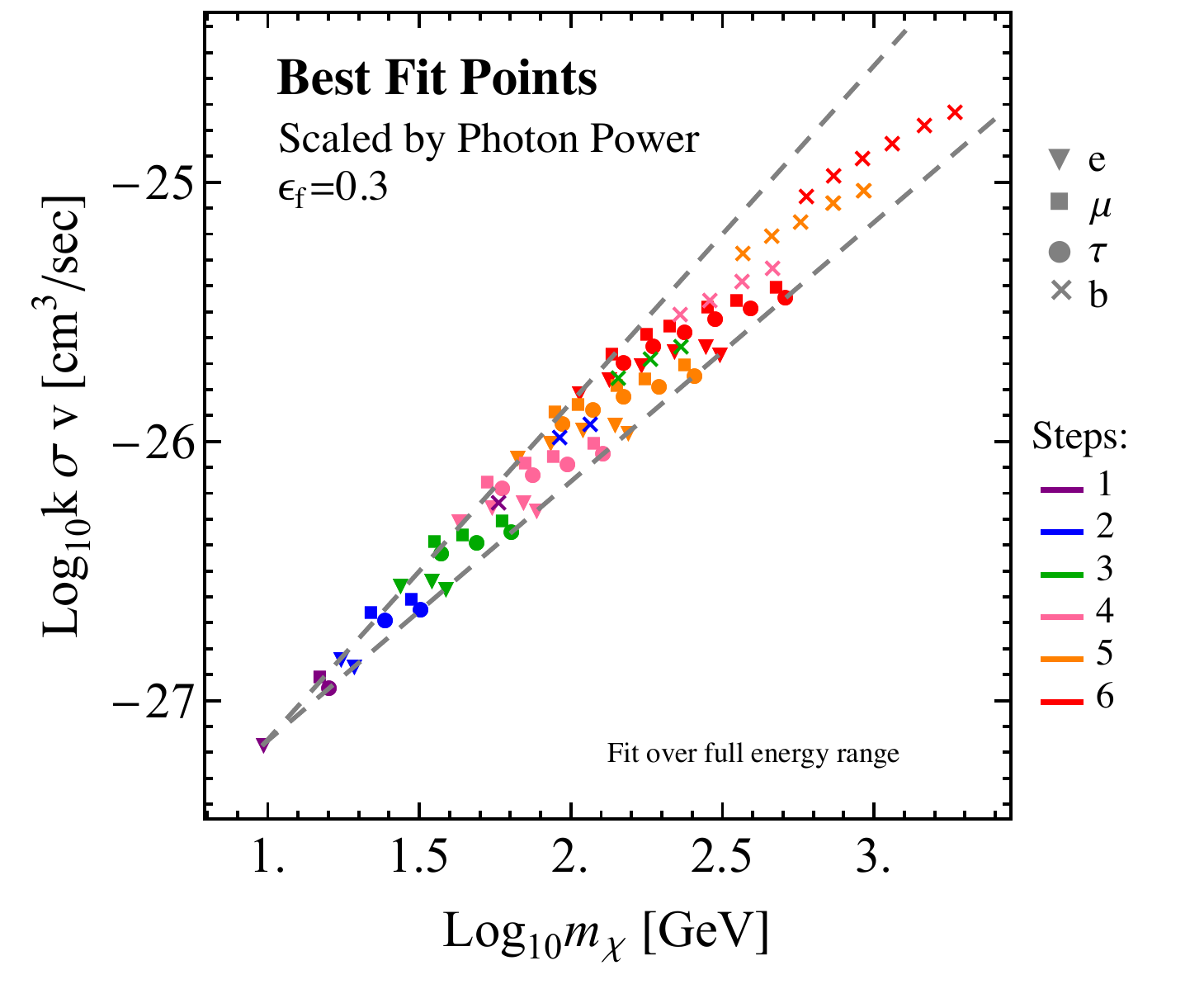}
\end{tabular}
\caption{\footnotesize{Colored points indicate the best fits for different numbers of hierarchical and degenerate cascade steps, and different final states, as in Fig.~\ref{fig:CombinedResults}. However, here we rescale the cross-section by the fraction of power into photons $k$ for each final state ($3.0 \times 10^{-3}$, $7.0 \times 10^{-3}$, $0.14$ and $0.26$ for electrons, muons, taus and $b$-quarks respectively). All final states then pick out the same region of $(m_\chi, k \sigma v)$ parameter space. The dashed lines indicate curves with $k \langle \sigma v \rangle \propto m_\chi$ and $k \langle \sigma v \rangle \propto m_\chi^{1.3}$, chosen to originate from the lowest-mass point studied; these curves approximately bound the full parameter space of interest (see text).}}
\label{fig:PhotonPowerScaled}
\end{figure}

The positions of the triangular regions in Fig.~\ref{fig:CombinedResults} largely reflect the differing branching ratios to photons (rather than other stable SM particles) for the different final states. For each of the direct annihilation (0-step) spectra, we can compute a factor $k$, defined as the total energy in photons (per annihilation) as a fraction of $m_1 = 2 m_\chi$. For example, direct annihilation/decay to $\gamma \gamma$ would have $k=1$. For the final states we consider, we find $k=3.0 \times 10^{-3}$, $7.0 \times 10^{-3}$, $0.14$ and $0.26$ for electrons, muons, taus and $b$-quarks respectively. Final states with smaller $k$ will naturally require higher cross-sections in order to fit the signal. In Fig.~\ref{fig:PhotonPowerScaled} we show the results of Fig.~\ref{fig:CombinedResults} replotted in terms of $k \langle \sigma v \rangle$ and $m_\chi$: we see that once this factor is taken into account, all channels pick out essentially the \emph{same} triangular region of parameter space, bounded by curves with $k \langle \sigma v \rangle \propto m_\chi$ and $k \langle \sigma v \rangle \propto m_\chi^{1.3}$.

\textit{Incorporating dark showers:} This concordance between the different final states suggests that dark shower models may be expected to also inhabit this region. For instance, the authors of \cite{Freytsis:2014sua} find a preferred cross-section of $8\times 10^{-27}$ cm$^3$/s for their $SU(2)_V$ model, with a roughly $35\%$ branching ratio into stable dark sector baryons (with other decay channels ending in photons), and a preferred mass of $\sim 10$ GeV. At first glance this suggests a somewhat higher value for $k \langle \sigma v \rangle$ than the lower tip of the triangular region identified in Fig.~\ref{fig:PhotonPowerScaled}. However, \cite{Freytsis:2014sua} fits to a different spectrum for the GCE excess (taken from \cite{2014arXiv1402.6703D}), without a systematic uncertainty estimate, and assumes a lower local DM density (0.3 GeV/cm$^3$ rather than 0.4 GeV/cm$^3$).\footnote{Private communication, Dean Robinson.} In our analysis, omitting systematic errors (or removing high-energy data points) raises the preferred cross-section by a factor of $\sim 2$ (Fig.~\ref{fig:Electron0p1StatsPlot}), and likewise lowering the local DM density from 0.4 to 0.3 GeV/cm$^3$ would raise the required cross-section by a factor of $\sim 2$; the lower tip of our triangular region would then reside at $m_\chi \sim 10$ GeV and $k \langle \sigma v \rangle \sim 3 \times 10^{-27}$ cm$^3$/s, which seems roughly consistent with \cite{Freytsis:2014sua}.

\subsection{Models with Vector Mediators}

Thus far we have considered models of multi-step cascades through scalar mediators. However models in which the hidden sector mediators include vector, fermion or pseudo-scalar particles are at least as equally well motivated (e.g. \cite{Pospelov:2007mp} or the dark shower example discussed above \cite{Freytsis:2014sua}). In the case of vector or fermionic mediators the simple recursion formula Eq.~\ref{eq:boosteq} will in general no longer hold, since the photon spectrum from the decay of mediators with spin need not in general be isotropic. The standard recursion formula will also break down if a decay is more than two-body, or if the decay is two-body but the decay products have different masses (although if the decay is strongly hierarchical the impact will be tiny), since these possibilities modify the Lorentz boost from the $\phi_i$ frame to the $\phi_{i+1}$ frame. Note this is different to having several possible decay chains with different branching ratios; in this case our analysis \emph{does} apply, and the final spectrum will simply be a linear combination of the spectra produced by the different decay chains.

Anisotropy of the photon spectrum is not in itself a sufficient condition for the recursion formula to break down. To modify the recursion, for some step $i$, the differential decay rate of $\phi_i$ must be a function of the angle $\theta$ between (1) the momenta of the decay products in the $\phi_i$ rest frame and (2) the boost direction from the $\phi_i$ rest frame to the $\phi_{i-1}$ rest frame. (Here we use $\phi_i$ to denote arbitrary mediators, independent of their spin.) Since the decays in the $\phi_i$ rest frame do not ``know'' about the $\phi_{i+1}$ frame, this sort of correlation is only possible if (1) the direction of the spin/polarization vector of the $\phi_i$ in its rest frame depends on the momentum with which it was produced in the $\phi_{i-1}$ rest frame, \emph{and} (2) the spectrum of the decay products of $\phi_i$ is a function of the angle between their momentum and the rest-frame spin/polarization vector of $\phi_i$. If only one of the two applies, averaging over the spin/polarization of $\phi_i$ will leave no $\theta$-dependence. However, both these properties will generically hold if $\phi_i$ is a vector: typically the decay of $\phi_{i-1}$ will prefer either longitudinally or transversely polarized vectors $\phi_i$, which will in turn decay with different angular distributions.

Let us consider the potential impact of such a $\theta$-dependence. For illustrative purposes, let us suppose that the \emph{photons} produced in the decays of $\phi_1$ (whether directly or by subsequent decays of the fermions) have essentially the same \emph{energy} spectrum as in the pure-scalar case, in the rest frame of the $\phi_1$. This assumption might fail if the spin of $\phi_1$ affects the correlations (if any) between the fermion spins, fermion momenta and photon momenta, but by making it we can isolate the impact of angular dependence in a single step of the cascade.

Consider a one step cascade $\chi \chi \rightarrow \phi_1 \phi_1$, $\phi_1 \rightarrow f \bar{f}$, where $\phi_1$ is a vector boson. Suppose the full spectrum of photons in the $\phi_1$ rest frame can be written as $\frac{dN}{dx_0} = f_{0}\left(y_0\right) dN/dx_0$, where $y_0 = \cos \theta_0$ and $dN/dx_0$ is the spectrum for the scalar mediator case $f_0 = 1$. Then the now familiar formula for the energy spectrum in the $\chi \chi$ center of mass frame is:
\be\begin{aligned}
\frac{dN_\gamma}{dx_1} &= 2 \int_{-1}^{1} d y_0 \int_{0}^1 dx_0 f_0\left(y_0\right) \frac{dN_\gamma}{dx_0}\\
& \delta\left( 2 x_1 - x_0 - y_0 x_0 \sqrt{1-\epsilon_1^2} \right) \\
&= 2 \int_{x_1}^1 \frac{dx_0}{x_0} f_0 \left( \frac{2x_1}{x_0} - 1 \right) \frac{dN_\gamma}{dx_0} + \mathcal{O}(\epsilon_1^2)\,.
\label{eq:Vectors}
\end{aligned}\ee
where we calculated the $y_0$ integral assuming $\epsilon_1 \ll 1$. Again we could extend this expression to an $n$-step cascade using the same formalism as in Appendix~\ref{app:boost}. The angular dependence at each step will in general be different depending on the model; we can parameterize this by specifying different functions $f_i\left(y_i\right)$ at each step. In the limit of small $\epsilon_i$ we find:
\be\begin{aligned}
 \frac{dN_\gamma}{dx_i} = 2 \int_{x_i}^1 \frac{dx_{i-1}}{x_{i-1}} f_{i-1}\left( \frac{2x_i}{x_{i-1}}-1 \right)\frac{dN}{dx_{i-1}} + \mathcal{O}(\epsilon_i^2)\,.
\label{eq:nStepVectors}
\end{aligned}\ee

\begin{figure}[t!]
\centering
\includegraphics[scale=0.593]{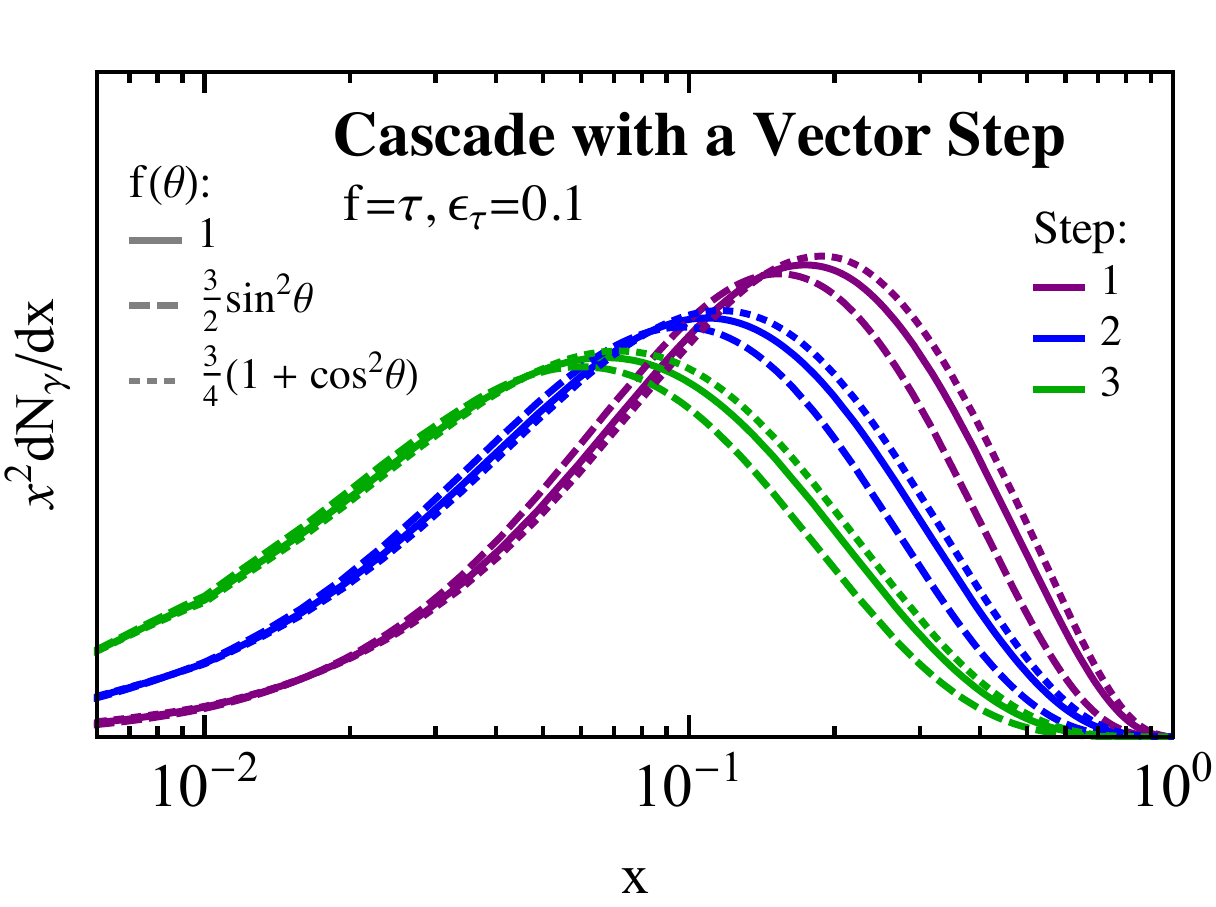}
\caption{\footnotesize{Spectrum for a 1-3 step cascade with a vector mediator in the final step of the cascade $\phi_2 \rightarrow V V$, $V \rightarrow  f \bar{f}$. We consider three separate cases: $f(\theta) = 1$, $(3/4)(1+\cos^2 \theta)$, and $(3/2)\sin^2\theta$. The first of these is equivalent to a cascade with only intermediate scalars (and hence isotropic decays), the others correspond to common angular dependences (see text).}}
\label{fig:vector}
\end{figure}

A detailed study of the impact of vector or fermionic mediators is beyond the scope of this paper; we leave it to future work. However, we can work out an explicit example motivated by the case where at the end of the cascade, a scalar/pseudoscalar resonance decays to two vectors which subsequently each decay into two fermions. This scenario has been studied in the context of Higgs decays \cite{Gao:2010qx}, furnishing results for a general resonance $X$ decaying to two identical vectors $VV$, which each in turn subsequently decay to $f \bar{f}$. (In our notation, the $V$ here would correspond to $\phi_1$ and $X$ to $\phi_{2}$.) The differential decay rate to fermions in this case is a linear combination of terms proportional to $\sin^2\theta$, $1 + \cos^2\theta$ and $\cos\theta$ (where $\theta$ is the angle defined above and in Appendix \ref{app:boost}), with coefficients depending on the axial and vector couplings of the fermions to the $V$, and the parity of the initial state $X$ \cite{Gao:2010qx}. In hierarchical decays of a scalar or pseudoscalar resonance to $VV$, where $V$ has vector (rather than axial vector) couplings to $f \bar{f}$, the dominant angular dependence is either $1 + \cos^2\theta$ or $\sin^2\theta$. For these specific (but common) angular dependences in the $\phi_1$ decay, we show the resulting changes to the photon spectrum in Fig. \ref{fig:vector}. The impact is modest, and so we expect our qualitative results should hold for more general cascades.

\section{Signals and Constraints}
\label{sec:signalsconstraints}

While we have remained agnostic regarding the choice of an actual model, we point out that any model with new light states in a dark sector that explains the GCE must also be consistent with the following experimental constraints:
\begin{itemize}
\item{Direct Detection: The coupling controlling $\sigma_{DD}$ must not be so large as to be in conflict with bounds from DM direct detection experiments \cite{Martin:2014sxa}.}
\item{Big Bang Nucleosynthesis (BBN): New light states must decay fast enough such that they do not spoil the predictions of BBN.}
\item{Collider constraints.}
\end{itemize}
These experimental constraints on a multi-step cascade will be very similar to those on a one-step cascade, with the key parameter being the coupling of the dark sector to the SM in both cases.

The simplest models that explain the GCE by direct DM annihilations to SM states are generally in conflict with direct detection bounds: the same coupling that must be small enough to avoid the LUX bound \cite{Akerib:2013tjd}, must also be large enough to explain the GCE with a thermal WIMP (note however that this conclusion is not inevitable; there are effective DM-SM couplings and simplified models that generically evade the bounds, e.g. \cite{Alves:2014yha, Berlin:2014tja}). As pointed out in \cite{Pospelov:2007mp,Martin:2014sxa,Abdullah:2014lla}, the addition of a dark sector with a single mediator allows for an explanation of the GCE while alleviating direct detection constraints. The reason is straightforward: any direct detection signal will be controlled by the coupling of the mediator to the SM, whereas the \emph{annihilation} rate is independent of this quantity, so the two can be tuned largely independently. We make this point more explicit in Appendix \ref{app:models}. Exactly the same property holds in models with expanded cascades, where the direct detection signal is controlled by the coupling between the dark sector and the SM; indeed, the direct detection signal may be suppressed even further if the coupling between the DM and the SM requires multiple mediators. If the couplings \emph{within} the dark sector are not highly suppressed, decays within the dark sector should in general proceed promptly (on timescales $\ll 1$ s), and so the constraint from BBN will primarily limit the coupling of the final mediator in the cascade to the SM. Accordingly, since it has been shown that for one-step cascades the constraint from BBN can be consistent with a null signal in direct detection experiments \cite{Martin:2014sxa}, the same should hold true for multi-step cascades (since in the multi-step case, the final step controls the coupling to the SM and hence provides the only relevant parameter for both BBN and direct detection). Collider bounds and limits from invisible decays of SM particles are also controlled by this final coupling, so can accordingly be dialled down in the same way as for one-step cascades, consistent with BBN bounds on the final coupling \cite{Martin:2014sxa}. A complex dark sector with multiple mediators could potentially give rise to interesting collider signatures (e.g. \cite{ArkaniHamed:2008qp, Baumgart:2009tn, Cheung:2009su}), but a detailed discussion is beyond the scope of this work.

\section{Conclusion}
\label{sec:conclusion}

We have laid out a general framework for characterizing the photon spectrum from multi-step decays within a secluded dark sector terminating in a decay to SM particles, and explored the ability of such a framework to produce the GeV gamma-ray excess observed in the central Milky Way. 

For any given SM final state, allowing multi-step decays expands the preferred region of $m_\chi-\langle \sigma v \rangle$ to a triangular region of parameter space, probed by cascades with different numbers of degenerate and hierarchical decays (where the decay products are slow-moving or relativistic, respectively), and bounded by curves with $\langle \sigma v \rangle \propto m_\chi$ and $\langle \sigma v \rangle \propto m_\chi^{1.3}$. Decays to different Standard model final steps correspond to different triangular regions in parameters space as shown in Fig.~\ref{fig:CombinedResults}. Large numbers of degenerate decays can raise the mass scale for the DM without bound, albeit at the cost of requiring a cross-section much higher than the thermal relic value and some degree of fine-tuning. Hierarchical decays broaden the photon spectrum, permitting a better fit to the data for SM final states that produce a sharply peaked photon spectrum; however, more than 4-5 hierarchical decays begin to reduce the quality of the fit even if the initial spectrum is very sharply peaked. In the absence of degenerate decays, the preferred mass range for the DM can then be constrained, and is consistently $\sim 20-150$ GeV across all channels; the corresponding cross-sections are close to the thermal relic value for tau and $b$-quark final states, and 1-2 orders of magnitude higher for $e$ and $\mu$ final states. Regardless of the final state, with the additional freedom of hierarchical decays the preferred spectrum tends to a similar shape, which can be approximated as the result of a cascade of 7-9 hierarchical decays terminating in a two-body $\gamma \gamma$ decay. We find that the best overall fits are still attained by DM annihilating to $b$-quarks (or other hadronic channels) with 0-2 hierarchical steps. 

Our preferred $\langle \sigma v \rangle-m_\chi$ regions are fairly insensitive to the details of the uncertainty analysis or the range of data points included. However, omitting high-energy data (above 10 GeV) substantially reduces the preferred number of hierarchical decay steps (from 4-5 to 2) for channels where the photon spectrum from direct annihilation is sharply peaked. There is currently disagreement between different analyses as to the high-energy photon spectrum associated with the excess; we do not take a position on this question, but note that its resolution may affect the range of dark-sector models that can provide viable explanations of the excess.

In this work we assumed that the directions of decay products in the rest frame of their progenitor are uncorrelated with the direction of the Lorentz boost to the rest frame of the previous progenitor particle in the sequence. Whilst always true for scalars, this may not hold for vector and fermionic mediators. We leave a more detailed discussion of concrete multi-step cascade models exploring these issues for future work.

\section*{Acknowledgements}

We would like to thank Douglas Finkbeiner, Andrew Larkoski, Ian Moult, Lina Necib, Matt Reece, Dean Robinson, Jessie Shelton, Jesse Thaler, Christoph Weniger, Wei Xue and Kathryn Zurek for helpful discussions and comments. This work is supported by the U.S. Department of Energy under grant Contract Number DE-SC00012567.

\appendix
\section{0-step Spectra}
\label{app:zerostep}
In order to calculate the photon spectrum, it is more straightforward to first determine the density of states according to:
\be\begin{aligned}
{\rm annihilations:}~\frac{1}{N_{\gamma}} \frac{dN_{\gamma}}{dE_{\gamma}} &= \frac{1}{\langle \sigma v \rangle} \frac{d \langle \sigma v \rangle}{d E_{\gamma}} \\
{\rm decays:}~\frac{1}{N_{\gamma}} \frac{dN_{\gamma}}{dE_{\gamma}} &= \frac{1}{\Gamma} \frac{d \Gamma}{d E_{\gamma}}
\label{eq:spectra}
\end{aligned}\ee
from which the spectrum can be easily backed out. Note that as pointed out in \cite{Fortin:2009rq}, if the cascade begins with a decay $\chi \to \phi_n \phi_n$, we will obtain an identical photon spectrum to the annihilation scenario, except the initial DM particle will be twice as heavy. This is the sense in which our results are readily transferred to the case of decaying DM. The key difference for the decaying case is the spatial morphology of the signal will generically require a line of sight integral over the DM density, rather than density squared as appears in the $J$-factor in Eq.~\ref{eq:Jfactor}. The observed spatial morphology of the GCE appears to disfavour decaying scenarios, which is why we do not mention them further here, although see \cite{Finkbeiner:2014sja} for a novel decay scenario that is distributed like density squared.

The result of Eq.~\ref{eq:spectra} is that in some circumstances it is possible to calculate various step cascades analytically. This approach is shown for several cases in \cite{Fortin:2009rq}. Yet in many cases - most notably those involving hadronic processes in their final states - analytic calculations are not feasible. For the present work we used a combination of analytic and numeric results depending on the final state employed. The details for each case is outlined below.

\subsection{Annihilations to $e^+ e^-$ }
The only contribution to the photon spectrum arises from FSR via the decay $\phi_1 \to e^+ e^- \gamma$. The spectrum in this case can be calculated analytically using Eq.~\ref{eq:spectra}, which was done in \cite{Mardon:2009rc} for the generic case of $\phi_1 \to f^+ f^- \gamma$. As pointed out there, when using the simple convolution formula Eq.~\ref{eq:boosteq}, consistency requires throwing away terms $\mathcal{O}(\epsilon_f^2)$ and higher, where $\epsilon_f = 2m_f/m_1$. Doing so they obtained the following expression for the spectrum that we include for completeness:
\be
\frac{dN_{\gamma}^{\rm FSR}}{dx_0} = \frac{\alpha_{\rm EM}}{\pi} \frac{1+(1-x_0)^2}{x_0} \left[ \ln \left( \frac{4(1-x_0)}{\epsilon_{f}^2} \right) -1 \right]\,.
\label{eq:FSR}
\ee
Note the $\ln$ term will dominate for small $\epsilon_f$, and the $-1$ is simply included to ensure consistency with the large hierarchies approximation. We confirmed that this spectrum is in agreement with the output from  \texttt{Pythia8} in the case of final state electrons. From here, by repeated use of the convolution formula it is possible to obtain completely analytic formula for the $n$-step cascade, which were used in our fits. For example, the first two steps are shown in \cite{Mardon:2009rc}.

\subsection{Annihilations to $\mu^+ \mu^-$ }
For final state muons, in addition to FSR, as pointed out in \cite{Mardon:2009rc} the radiative decay of the muon $\mu \rightarrow e \bar{\nu}_e \nu_{\mu} \gamma$ will meaningfully contribute to the photon spectrum. This decay was calculated in \cite{Kuno:1999jp}, and again for completeness we include it here as it was presented in \cite{Mardon:2009rc}:
\be
\frac{dN_{\mu\to\gamma}}{dx_{-1}} = \frac{\alpha_{\rm EM}}{3\pi} \frac{1}{x_{-1}} \left( T_{-1}(x_{-1}) \ln \frac{1}{r} + U_{-1}(x_{-1}) \right)\,,
\ee
where $r=m_e^2/m_{\mu}^2$ and
\be\begin{aligned}
T_{-1}(x) =& (1-x)(3-2x+4x^2 - 2x^3) \\
U_{-1}(x) =& (1-x) \left( -\frac{17}{2} + \frac{23}{6} x - \frac{101}{12} x^2 + \frac{55}{12} x^3 \right. \\
&\left.+ (3-2x + 4x^2 - 2x^3) \ln(1-x) \right)
\end{aligned}\ee
Note the subscript $-1$ here is used to remind us this is the spectrum calculated in the rest frame of the muon. To then obtain the 0-step cascade we would have to apply Eq.~\ref{eq:boosteq} once, assuming $\epsilon_{\mu} = 2m_{\mu}/m_1 \ll 1$, and then combine this with the FSR spectrum in Eq.~\ref{eq:FSR}.

\subsection{Annihilations to $\tau^+ \tau^-$ }
For the case of final state taus, FSR will now be a subdominant contribution. Instead the spectrum will have a much larger contribution from leptonic and semi-leptonic tau decays: $\tau^- \rightarrow \nu_\tau  l^- \bar{\nu_l}$ and $\nu_\tau d \bar{u}$. The quarks will then hadronize (dominantly to pions) which will result in large contributions to the photon spectrum. We simulated this final state in  \texttt{Pythia8} to generate an initial spectrum, to which we could then apply the convolution formula.

\subsection{Annihilations to $b \bar{b}$}
Much like for taus, in the case of final state $b$-quarks FSR is a subdominant contribution, and instead the spectrum is largely determined by hadronic processes. As such we again utilize  \texttt{Pythia8} to obtain the initial spectrum.

\section{Kinematics of a Multi-step Cascade}
\label{app:boost}

As already emphasized the utility of the small $\epsilon_i=2m_i/m_{i+1}$ - or large hierarchies - approximation is threefold: 
\begin{enumerate}
\item It simplifies calculations in that we can use Eq.~\ref{eq:boosteq}, rather than the general formula we display below; 
\item More importantly it allows us to describe a cascade using just the identity of the final state $f$, the value of $\epsilon_f$, and the number of steps $n$, in contrast to the many possible parameters of the generic case; 
\item Despite the simplifications afforded, results in this framework can be used to estimate the results even for general $\epsilon_i$, as described in Sec.~\ref{sec:generalcascade}.
\end{enumerate}
 In this appendix we show how the kinematics of scalar cascade decays lead to an expression for the $n$-step spectrum in terms of the $(n-1)$-step result. In addition we outline how Eq.~\ref{eq:boosteq} emerges in the small $\epsilon$ limit, with error $\mathcal{O}(\epsilon_i^2)$, as well as how the transition to the degenerate case as $\epsilon \to 1$ occurs.

Our starting point is the 0-step spectrum $dN_\gamma / dx_0$ where $x_0 = 2 E_0 / m_1$ and $E_0$ is the photon energy in the rest frame of $\phi_1$. This results from the process $\phi_1 \to \gamma X$, where the identity of $X$ depends on the final state considered. From here we want to calculate $dN_\gamma / dx_1$ - the spectrum from a cascade that includes $\phi_2 \to \phi_1 \phi_1$ and so is one step longer - where $x_1 = 2 E_1/m_2$ and $E_1$ is the photon energy in the $\phi_2$ rest frame. If we assume isotropic scalar decays, then we can obtain this by simply integrating the 0-step result over all allowed energies and emission angles:
\be\begin{aligned}
\frac{dN_{\gamma}}{dx_1} = & 2 \int_{-1}^1 d \cos \theta \int_{0}^1 dx_{0} \frac{dN_{\gamma}}{dx_{0}} \\
&\delta \left( 2x_1 - x_0 - \cos \theta x_0 \sqrt{1-\epsilon_1^2} \right)\,,
\end{aligned}\ee
where $\theta$ is defined as the angle between the photon momentum and the $\phi_1$ boost axis as it is measured in the $\phi_1$ rest frame. The limits of integration $0 \leq x_0 \leq 1$ reflect the fact that the photon energy in the $\phi_1$ rest frame can be arbitrarily soft on the one side, and on the other it can have an energy at most half the mass of the initial particle, $m_1/2$ here. The $\delta$ function is simply enforcing how the photon energy changes when we move from the $\phi_1$ to the $\phi_2$ rest frame, i.e. from $E_0$ to $E_1$. It also sets the kinematic range for $x_1$, which is:
\be
0 \leq x_1 \leq \frac{1}{2} \left(1+ \sqrt{1-\epsilon_1^2}\right)\,.
\label{x0Range}
\ee
Now if we then use the $\delta$ function to perform the angular integral, the one step spectrum reduces to:
\be
\frac{dN_\gamma}{dx_1} = 2 \int_{t_{1,{\rm min}}}^{t_{1,{\rm max}}} \frac{dx_{0}}{x_0 \sqrt{1- \epsilon_{1}^2}} \frac{dN_\gamma}{dx_{0}}~\,,
\label{eq:fullepsilonstep1}
\ee
where we have introduced:
\be\begin{aligned}
t_{1,{\rm max}} &= \min \left[ 1,\, \frac{2x_1}{\epsilon_1^2} \left( 1 + \sqrt{1-\epsilon_1^2} \right) \right] \\
t_{1,{\rm min}} &= \frac{2 x_1}{\epsilon_1^2}\left( 1 - \sqrt{1-\epsilon_1^2} \right)
\label{eq:1stepminmax}
\end{aligned}\ee
The maximum here is either set by the maximum physical value of $x_0$, which is $1$, or alternatively by where the $\delta$ function loses support. We can then repeat this process to recursively obtain the $i$th order spectrum from the $(i-1)$th order result. Explicitly we find:
\be
\frac{dN_\gamma}{dx_i} = 2 \int_{t_{i,{\rm min}}}^{t_{i,{\rm max}}} \frac{dx_{i-1}}{x_{i-1} \sqrt{1- \epsilon_{i}^2}} \frac{dN_\gamma}{dx_{i-1}}~\,,
\label{eq:fullepsilon}
\ee
where we have defined:
\be\begin{aligned}
t_{i,{\rm max}} &=  \min \left[ \frac{1}{2^{i-1}} \prod_{k=1}^{i-1} \left( 1 + \sqrt{1- \epsilon_k^2} \right),\right. \\
&\left.\;\;\;\;\;\;\;\;\;\;\;\;\;\frac{2x_i}{\epsilon_i^2} \left( 1+\sqrt{1-\epsilon_i^2}\right) \right] \\
t_{i,{\rm min}} &= \frac{2 x_i}{\epsilon_i^2}\left( 1 - \sqrt{1-\epsilon_i^2} \right)
\label{eq:nstepximin}
\end{aligned}\ee
and now the kinematic range of $x_i$ is
\be
0 \leq x_i \leq \frac{1}{2^i} \prod_{k=1}^i \left( 1 + \sqrt{1- \epsilon_k^2} \right)\,.
\label{eq:xnRange}
\ee
With the exact result of Eq.~\ref{eq:fullepsilon}, we can now see that in the small $\epsilon$ limit the result reduces to Eq.~\ref{eq:boosteq} with corrections at most of order $\epsilon^2$, as claimed. The exact result also captures an additional feature that the large hierarchies result does not: the emergence of a degenerate step in the cascade as $\epsilon_i\to 1$ for some $i$. As discussed in Sec.~\ref{sec:generalcascade}, when this occurs, just from the kinematics we can see that the $(i+1)$-step result will reduce to the $i$-step spectrum, but shifted in energy and normalisation. Starting with Eq.~\ref{eq:fullepsilon}, setting $1-\epsilon_i^2 \equiv z$ and then taking $z \to 0$ it is straightforward to confirm that the exact result also reproduces this behaviour.

As discussed in Sec.~\ref{sec:generalcascade}, there should be a smooth interpolation between the two extreme cases of $\epsilon_i=0$ and $\epsilon_i=1$, and using Eq.~\ref{eq:fullepsilon} we can demonstrate that indeed there is. This is shown in Fig.~\ref{fig:fullEpsilon0p1}, where we take the case of a 1-step cascade for final state taus with $\epsilon_{\tau}=0.1$. We plot the two extreme cases and show how intermediate $\epsilon$ transition between these by plotting five values: $0.3$, $0.5$, $0.7$, $0.9$ and $0.99$. Note that as claimed earlier, the transition is roughly quadratic in $\epsilon$; for small and intermediate values of $\epsilon$, the result is well approximated by the $\epsilon=0$ result, again highlighting the utility of the large hierarchies approximation.

\section{Model-Building Considerations}
\label{app:models}

\subsection{A Simple Model}

Let us extend the usual Higgs Portal \cite{Patt:2006fw, MarchRussell:2008yu} model to include a rich dark sector with $n$ scalar mediators and a set of $n$ $\mathbb{Z}_2$ symmetries.\footnote{A more complex symmetry structure could allow off-diagonal couplings between the scalars and the Higgs, with potentially rich observational signatures. We thank Jessie Shelton for this observation.} This will serve as an illustrative example of how different observable signatures depend on different model parameters, as discussed in the main text.

Consider the potential:
\be\begin{aligned}
&V\left(\chi, \phi_1, H\right )  = V_\chi + V_H + c_k \phi_1^2 |H|^2 \\
&+ \sum_{i = 1}^{n}\left(\frac{\lambda_{4,i}}{2} \chi^2 \phi_{i}^2 -\frac{1}{2}m_{i}^2 \phi_i^2\right) + \sum_{i,j = 1}^{n}  \frac{\lambda_{ij}}{4 !}\phi_i^2 \phi_j^2\,, 
\label{eq:HiggsPortal}
\end{aligned}\ee
Here $V_\chi$ and $V_H$ contain the usual mass and quartic terms for the DM and Higgs fields. As discussed previously it is reasonable that the dark sector is secluded such that the dominant portal coupling is $c_k \phi_1^2 |H|^2$. Upon electroweak and $\mathbb{Z}_2$ symmetry breaking the $\lambda_{4,i}$ couplings allow annihilations $\chi \chi \rightarrow \phi_i \phi_i$. We assume that DM annihilates preferentially to the heaviest mediator through $\lambda_{4,n} \chi^2 \phi_n^2$. So it is $\lambda_{4,n}$ that dominantly controls the thermal annihilation cross-section and therefore the DM relic abundance $\Omega_\chi h^2 \sim 0.11$. The dark sector quartic term will generate interactions of the form $\lambda_{ij}  \langle \phi_i \rangle \phi_i \phi_{j}^2 $, allowing the mediators to cascade decay in the dark sector. Additionally the Higgs Portal interaction will generate a mixing between $\phi_1$ and the Higgs. The end result will be a dark cascade ending in the $c_k$ suppressed decay $\phi_1 \rightarrow f \bar{f}$, with a subsequent photon spectrum that can be fit to the GCE. 

While the thermal relic cross-section depends on $\lambda_{4,n}$, the direct detection cross-section will also depend on the portal coupling $c_k$. This additional small parameter gives us the needed freedom to explain the GCE while alleviating constraints from direct detection. Additionally we point out that the size of the couplings $\lambda_{ij}$ will need to be large enough such that  decays of the new light states occur before BBN. Given the number of new free parameters, this setup should not be difficult to construct. Finally we point out that the Higgs Portal interaction also contains a coupling which leads to the decay $h \rightarrow \phi_1 \phi_1$. Invisible Higgs decay is constrained by collider searches which impose an upper bound of about $c_k \lesssim 10^{-2}$ \cite{Martin:2014sxa}.

\subsection{The Sommerfeld Enhancement}

We have seen that the preferred cross-section steadily increases with the number of steps in the cascade, moving away from the thermal relic value that is favored for the direct case. This increased cross-section is also accompanied by an increase in the preferred mass scale for the DM (indeed, the requirement for a larger cross-section is largely driven by the reduced number density of heavier DM). In the presence of a mediator much lighter than the DM, exchange of such a mediator could enhance the present-day annihilation cross-section via the Sommerfeld enhancement (e.g. \cite{Sommerfeld:1938,Hisano:2003ec, Hisano:2004ds,ArkaniHamed:2008qn, Pospelov:2008jd}), naturally leading to an apparently larger-than-thermal annihilation signal.

However, there are some obstacles to such an interpretation, at least in the simple case we have studied where the particles involved in the cascade are all scalars. For the case of fermionic DM coupled to a light scalar or vector of mass $m_\phi$ with coupling $\alpha_D$, the Sommerfeld enhancement at low velocity is parametrically given by $m_\phi/\alpha_D m_{\chi}$. A large enhancement thus requires $\alpha_D \gtrsim m_\phi / m_\chi$. In order to obtain the correct relic density, we typically require $\alpha_D$ to be $\mathcal{O}(0.01)$, and so a significant Sommerfeld enhancement would require the \emph{first} step in the cascade to involve a mass gap of two orders of magnitude. This may be plausible for the electron and even muon channels, but is challenging for final states involving heavier particles such as taus and $b$-quarks; if the mediator is heavy enough to decay to these particles, the required DM mass becomes much too large to fit the GCE even for a one-step cascade, and adding more hierarchical steps only exacerbates the self-consistency issue (as discussed in Secs.~\ref{sec:methods}-\ref{sec:results}).

Furthermore, if the DM is a fermion, its annihilation into scalars is generically $p$-wave suppressed, making it difficult to obtain a large enough cross-section to obtain the GCE. If instead the DM is a heavy (singlet) scalar, the simplest way to couple it to the light scalar to which it annihilates is an interaction of the form $\mathcal{L}_\mathrm{quartic} = \frac{\lambda_4}{2} \chi^2 \phi_{n}^2$. When the light scalar obtains a vacuum expectation value, this gives rise to an interaction of the form $\lambda_4 \langle \phi_n \rangle \phi_n \chi^2$, and repeated exchanges of the light scalar $\phi_n$ can give rise to enhanced annihilation. However, assuming $\langle \phi_n \rangle \sim m_n$, the size of the coupling is suppressed by the small mass of the light scalar, even as its range is enhanced. Accordingly, a large enhancement to annihilation is not expected, at least in this simple scenario.

As discussed in Sec.~\ref{sec:generalcascade}, our results can be extended to cascades including particles other than scalars, in which these later issues do not arise; for example, in the axion portal \cite{Nomura:2008ru}, two-step cascades occur through $\chi \chi \rightarrow s a$, $s \rightarrow a a$, $a \rightarrow f \bar{f}$, where $s$ is a dark scalar and $a$ a dark pseudoscalar. This annihilation channel is $s$-wave and can be Sommerfeld-enhanced by exchange of the $s$. However, the first difficulty described above may still apply, with the large hierarchy between the $\chi$ and $s$ potentially implying a DM mass too large to easily fit the GCE.

\bibliography{cascades}

\end{document}